\documentclass[prb,showpacs,twocolumn,amsmath,amssymb]{revtex4}
\usepackage{graphicx}% Include figure files
\usepackage{dcolumn}% Align table columns on decimal point
\usepackage{bm}% bold math
\begin{document}
\title{Particle Velocity Fluctuations in Steady State Sedimentation: Stratification Controlled  Correlations.}
\author{P.N. Segr{\` e} and J.E. Davidheiser}
\address{Department of  Physics, Emory
University, Atlanta, GA 30322}
\begin{abstract}
The structure and dynamics of  steady state sedimentation of
semi-concentrated ($\phi=0.10$)  monodisperse spheres are studied in
liquid fluidized beds. Laser turbidity and particle imaging methods
are used to measure the particle velocity fluctuations and the
steady state concentration profiles.  Using a wide range of particle
and system sizes, we find that the measured gradients
      $\nabla \phi$, the fluctuation magnitudes $\sigma_v$, and their spatial      correlation lengths $\xi$, are not uniform in the columns - they all show strongly $z-$dependent profiles.
      These profiles also display a  scaling
      in which results from different particle sizes
     collapse together when plotted in the forms $-a\nabla \phi(z)$,
     $\xi(z)/a$, and $\sigma_v(z)/v_p$,
     demonstrating the universality of the particle dynamics and structure
     in steady state sedimentation. Our results are also used to test a
     recently proposed model for the correlation lengths $\xi(z)$ in
     terms of the  concentration stratification $\nabla \phi(z)$
     [P.J. Mucha and M.P. Brenner, Phys. Fluids {\bf 15}, 1305 (2003)],
     $\xi(z)=c_0 a[\phi
S(\phi)]^{1/5}[-a\nabla\phi(z)]^{-2/5}$.
We  find that the correlation lengths predicted by this  model
are in very good agreement with our measured values, showing that
the origin of the fluctuation length $\xi$ lies with the
concentration stratification $\nabla \phi$.
\end{abstract}
%%% ----------------------------------------------------------------------
\maketitle
%%% ----------------------------------------------------------------------

\section{INTRODUCTION}The slow sedimentation of a collection
of non-Brownian spheres in liquids is a fundamental problem in
physics and is of importance in numerous chemical industries
\cite{hinch}. Despite its apparent simplicity, the complexities of
hydrodynamic interactions between particles  are still vigorously
debated \cite{candl,koch,alex,penger,mike1}. Much of the debate
surrounds the origin of the characteristic  size of velocity
fluctuations, $\xi\sim 11a\phi^{-1/3}$, found in sedimentation
experiments \cite{segreprl,nature} (Here $a$ is the particle radius
and $\phi$ is the volume fraction). This lengthscale is typically
smaller than the cell dimensions and
 plays a central role in sedimentation dynamics.
The magnitude of the particle velocity fluctuations $\sigma_v$
depends directly on $\xi$.
   In dilute suspensions,  Poisson statistics dictates that in a
region of size $\xi$, the typical concentration fluctuation is
$\sigma_\phi=\sqrt{\phi a^3/\xi^3}$; its velocity is determined by
balancing its buoyant weight, $\sigma_\phi \Delta\rho g\xi^3$, with
its Stokes drag, $6\pi\eta \xi \sigma_v$, giving $\sigma_v\propto
\sqrt{\phi \xi/a}$.\cite{segreprl,bruce} Similarly, particle
diffusion coefficients, $D\sim\xi \sigma_v$, also  depend directly
on $\xi$.\cite{nature} Due to its central role, an understanding of
the origin of the fluctuation correlation length is essential for
even a basic description of sedimentation.

Recent theories \cite{luke,shang,muchaPOF,jfm,shaqfeh} have taken a
new approach to understanding the lengthscale of velocity
fluctuations by considering the possible effects of a small
stratification in mean column concentration, $\nabla\phi$.  Mucha
and Brenner \cite{muchaPOF,jfm} argue that because velocity
fluctuations of size $l$ are driven by buoyancy forces acting on
local concentration fluctuations $\sigma_\phi$, when there is a
concentration variation across the fluctuating region due to the
stratification $l\nabla \phi$ that is  of order the fluctuation
itself, $\sigma_\phi=\sqrt{\phi a^3/l^3}$, then the buoyancy
mismatch of the fluctuating region is reduced, and motions are
suppressed. The largest region for which motion occurs then is of
order $l=-\sigma_\phi/\nabla \phi$, or
 \begin{equation}
l/a=c_0 [\phi S(\phi)]^{1/5}[-a\nabla\phi]^{-2/5}  , \label{eq:bm}
\end{equation}
which relates the correlation size $l$ to the concentration
stratification $\nabla \phi$. Here, the static structure factor
$S(\phi)$ accounts for excluded volume affects in non-dilute samples
\cite{nature}. Results in support of this model \cite{muchaPOF},
using numerical simulations, have focussed on the
  diffusion coefficients at the transition
region near the top particle interface. The generality of the Mucha
and Brenner (MB) model, however, suggests  it is not confined solely
to interfacial properties, and may provide a complete description
for the fundamental length $\xi$ over the entire sedimenting column.

The most direct test of the MB model, as performed in this article,
would be a direct comparison between measured fluctuation lengths
$\xi$, and the calculated stratification lengths $l$, over all
heights in a sedimenting
 particle column. This test has yet to be done, principally due to a lack of
 data for both   $\xi(z)$ and $\nabla\phi(z)$.  Moreover, the few
available data for $\nabla\phi$ far from the interface, using
simulations, yield seemingly contradictory results; Ladd
\cite{tony1} found negligibly small gradients, but Mucha and Brenner
\cite{muchaPOF} found much larger ones. We note that since Mucha and
Brenner considered larger systems than Ladd, this contradiction may
be resolved if gradients have an as yet unidentified system size
dependency.

In this article we examine steady state sedimentation using a wide
range of particle and system sizes. The samples have the same column
height, $H$, and the same mean concentration, $\phi_0=10.0\%$, but
different particle sizes $a$, which range over a factor of
 $\sim 4.5$ so that the system size $N$ ($\propto a^3$)
 varies by almost two orders of magnitude.
 We first show that all of the samples show some
 degree of stratification. The magnitude of the
 stratification increases with (normalized)
cell size until its average becomes of order
$\overline{\nabla \phi}\sim \phi_0/H$.
 We then show that there is a scaling regime in
 which the (height-dependent) concentration
gradients, the velocity correlation lengths, and the velocity
fluctuation magnitudes all  collapse onto master curves when scaled
as $-a\nabla \phi(z)$, $\xi(z)/a$, and $\sigma_v(z)/v_p$. All of our
data are consistent with the MB model. We find that the
stratification lengths $l(z)$ calculated from Eq. (\ref{eq:bm})
agree with our measured values $\xi(z)$ using a single fit constant,
$c_0 =0.82$, for three different samples in the scaling regime, and
a slightly lower value, $c_0=0.70$, in the non-scaling regime.
 These findings validate the physical picture behind the MB model,
 and show that the origin of the fluctuation length $\xi$ lies
 with the concentration stratification $\nabla \phi$.
\begin{figure}\includegraphics[width=70mm]{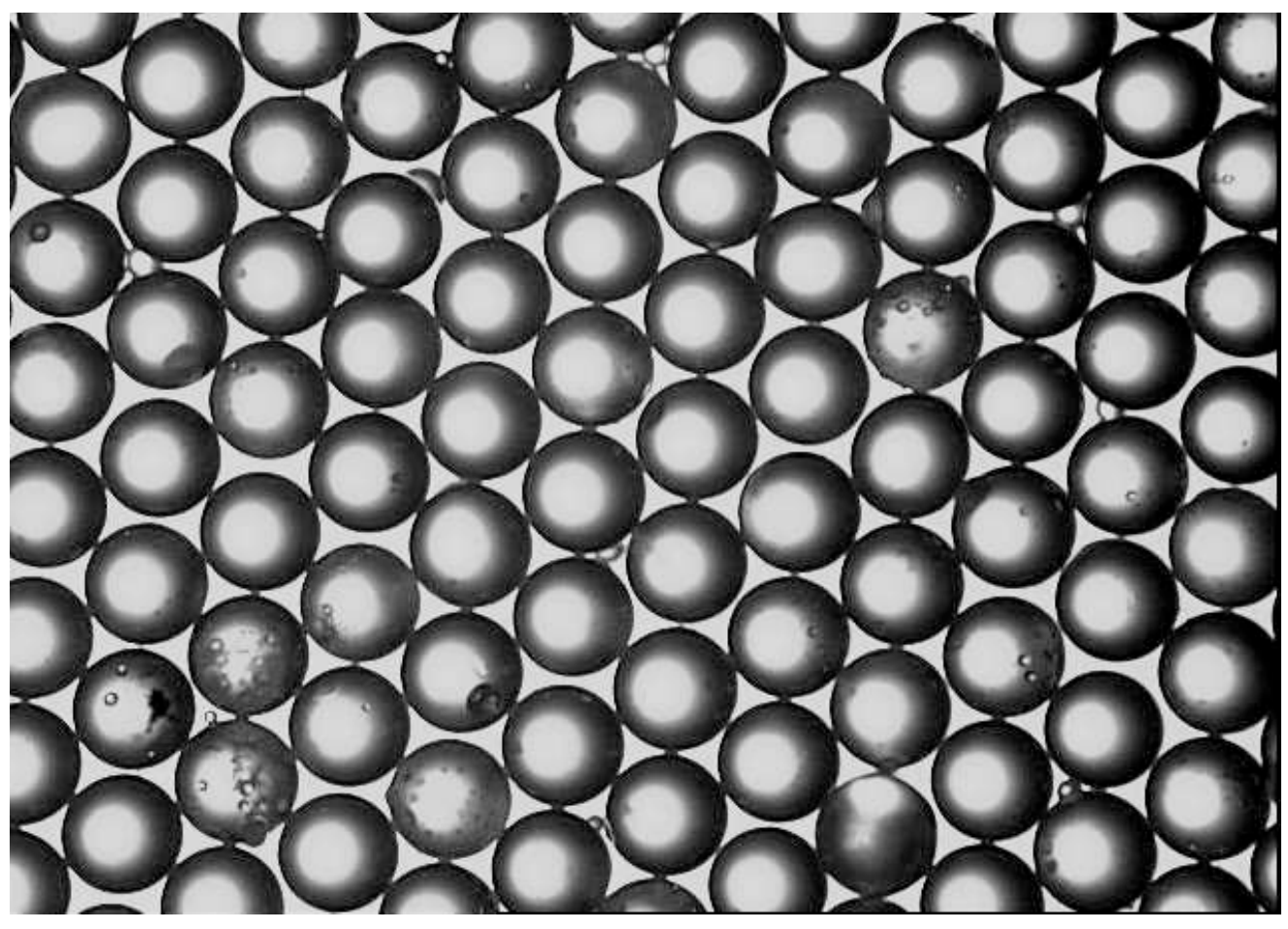}
\caption{Photograph of a random sampling of $a=154.5 \;\mu$m spheres
resting on a glass slide. } \label{fig:particle}
\end{figure}

Our choice of an experimental arrangement to study $\nabla\phi$ is
guided by the recent observations of  persistently evolving
gradients during sedimentation \cite{shang,evolution,maxey}.
    The evolution occurs because,
    starting from a well mixed sample ($\nabla \phi=0$), it takes some time for the stratification ($\nabla\phi\neq0$) to become
    established. During this evolution,  the column is falling, and often
   steady state is never reached before all the particles
lie on the bottom  \cite{shang}. To ensure we reach a steady state,
we conduct experiments in a fluidized bed apparatus, where liquid is
pumped upwards to
    balance the settling of the particles \cite{salin,fbprl}.
The flow balancing ensures that the particle column never collapses,
and a steady state is eventually established.
\section{EXPERIMENT DESCRIPTION}
\subsection{Particles and Fluids}
\begin{figure} \includegraphics[width=80mm]{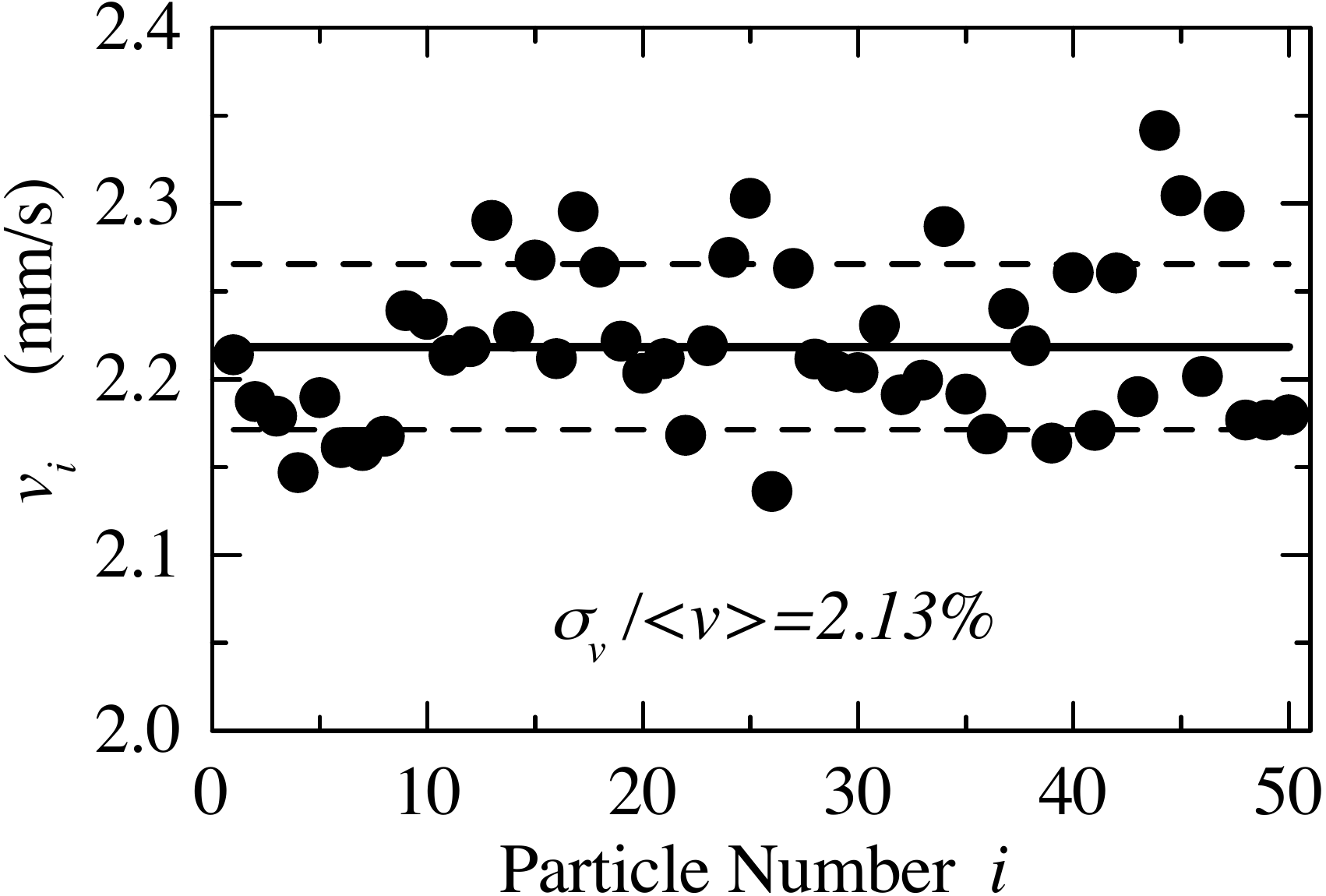}
\caption{The settling velocities of $50$ isolated spheres
falling in a water/glycerol mixture. The solid line is the average
settling rate, $v=2.218$ mm/s, and the dashed lines the standard
deviation, $\sigma_v=0.047$ mm/s. The polydispersity in velocity is
$\sigma_v/v=2.13 \%$, corresponding to a particle size polydispersity of
$\sigma_a/a=1.07 \%$.} \label{fig:partdist}
\end{figure}

We use spherical glass
    beads of mean radii $a=54.5, 103.5, 154.5$, and $245$ $\mu$m.
     The beads have been specially filtered by the manufacturer (Mo-Sci corp.)
    to be of extremely low polydispersity in size. To demonstrate this, we show in
    Fig. \ref{fig:particle}  a micrograph of a collection of the
    $a=154.5\; \mu$m beads.  The particles lay on a flat plate and
    through a gentle shaking quickly organize into highly ordered
    forms, indicative of low levels of polydispersity $\sigma_a$.
   To accurately determine $\sigma_a$,
   for each set of particles we measure the
   variation in settling velocities of $50$ individual
    spheres. Fig. \ref{fig:partdist} shows the results for the $a=245\;
    \mu$m spheres. The standard deviation of the
    individual particle settling rates is indeed
     extremely low, $\sigma_v/v=2.13 \%$.
    Using the Stokes formula\cite{hinch}, $v_0=(2/9) \Delta \rho ga^2/\eta$,
    and assuming the particles are all of equal
    density, the variation in settling rates correspond to a
    variation in particle size of $\sigma_a/a=1.07 \;\%$. This procedure
    is repeated for all $4$ particle sizes, with the results
    listed in Table I. In all cases,
    the measured value of the size polydispersity
    $\sigma_a/a$ is less than $1.7\%$.

    The  glass  beads are
    dispersed in viscous solutions of glycerol
    and water chosen so that inertial forces during fluidization
    are negligible. The particle Reynolds numbers, listed in Table
    I, are all of order $Re=v_0a \rho/\eta \approx 5 \times 10^{-3}\ll
    1$. Additionally, particle motions always occur at
    very high Peclet numbers ($Pe\approx 10^9$),
     so that Brownian diffusion is negligible.
    The temperature is at the ambient value $T=21 ^o$C.

    \begin{table}
\caption{Particle, fluid, and fluidized bed properties. $a$ is the
average particle radius, $\sigma_a/a$ the measured size
polydispersity, $\eta$ the pure solvent viscosity, $v_0$ the Stokes
settling velocity, $Re$ the particle Reynolds number, $H$ the total
height of the fluidized particle column, and $\phi_0$ the average
volume fraction of the fluidized bed.}
\begin{tabular}{ccccccc}
 $a (\mu m)$ & $\sigma_a/a$ &$\eta$ (cp)&$v_0 (mm/s)$&$Re$&$H$ (cm)&
 $\phi_0$\\
\tableline
$54.5$&$0.016$&$17$&$0.548$&$0.002$& $18.5$&$0.10$ \\
$103.5$&$0.015$&$27$&$1.21$&$0.005$&$18.5$&$0.10$\\
$154.5$&$0.012$&$54$&$1.34$&$0.004$& $18.5$&$0.10$ \\
$245$&$0.011$&$87$&$2.10$&$0.006$& $18.5$&$0.10$ \\
\tableline \tableline
\end{tabular}
\end{table}
\subsection{Fluidized Bed}
\begin{figure} \includegraphics[width=70mm]{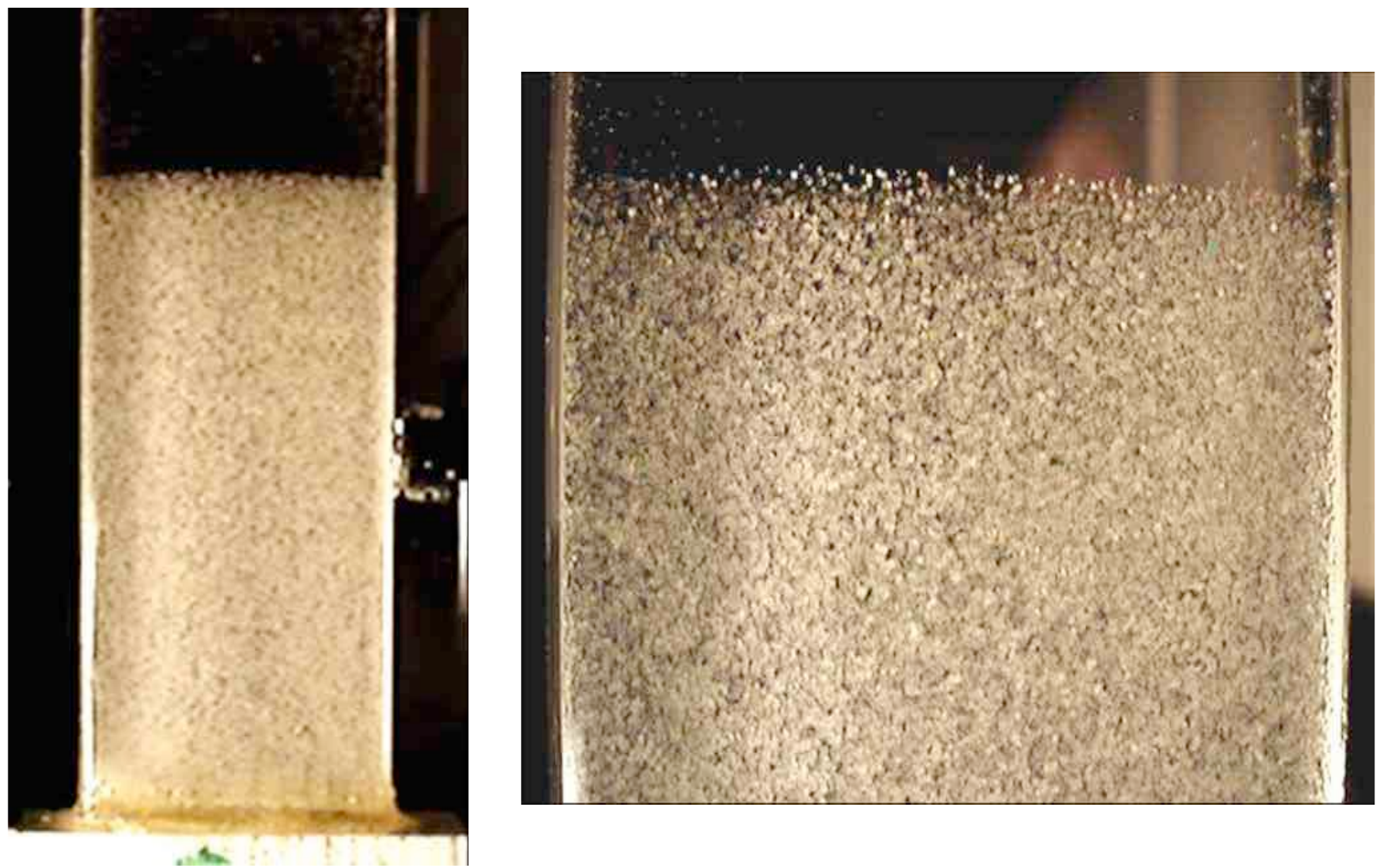}
\caption{Photographs of a fluidized particle bed comprised of $D=490
\;\mu$m spheres. LHS - whole cell. RHS - closeup of the top of the
column.} \label{fig:photobed}
\end{figure}
The fluidized bed, shown in Fig. \ref{fig:photobed}, consists  of a
fluid and particle filled glass cell at the bottom of which fluid is
pumped upwards to counteract the particle settling and {\it
fluidize} the particles.
    The sample cell is a rectangular glass tube of dimensions
    $T\times W\times H = 8\times 80\times 305$ mm.
    The overflow liquid at the top of the cell recirculates back
into the pump, forming a closed loop. To enable a uniform flow into
the cell, a $2$ cm thick nylon mesh is packed with $0.5$ mm diameter
beads and glued across the entrance to the cell at the bottom.

In the experiments described below, we use this setup to fluidize
four different particle size samples, as listed in Table I. To
examine the specific effects of particle size, all four samples
share the same average particle concentration, $\phi_0 =0.10$, and
same total height of the particle column $H$. With the pump off, the
spheres in all $4$ samples form a sediment at the bottom $\sim 2.9$
cm tall. When the pump is on, the particles expand upward, filling a
region above the bottom up to a height dependent upon the pumped
fluid velocity $v_{p}$ (see Fig. \ref{fig:photobed}). For all $4$
samples, we set $v_{p}$ to a value that expands the particle column
to a total height $H\sim 18.5$ cm, so that the average volume
fraction is approximately the same for all of the samples,
$\phi_0=\phi_{RCP}*(2.9/18.5)\sim 0.10$, where $\phi_{RCP}=0.638$ is
the assumed value for the sediment at the random close packing
density.
\subsection{PIV Imaging System: Velocity Flow Maps} \label{sec:piv}
\begin{figure} \includegraphics[width=75mm]{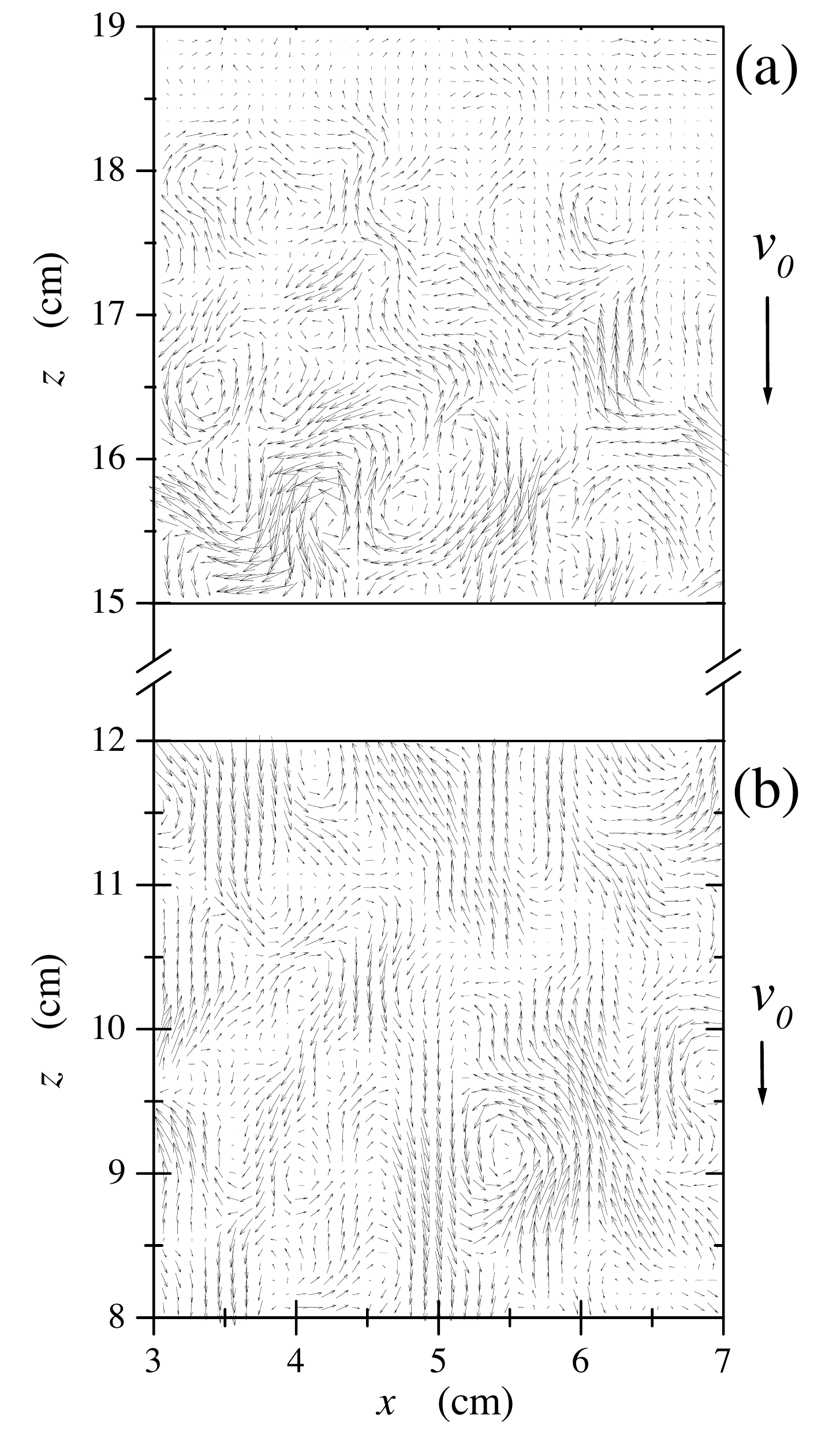}
\caption{Velocity vector maps of a stable fluidized bed of average concentration $\phi_0=0.100$. Figures 1 (a) and (b) correspond to
respective positions near to the top and near to the middle of the
particle column. The single arrow on the right gives the
corresponding scale of the Stokes settling velocity $v_0$. Note that
the velocity scale in (a) is magnified relative to (b) by a factor
of $2$ for clarity.} \label{fig:piv}
\end{figure}Particle velocities are measured using the technique of particle
image velocimetry (PIV) \cite{adrian}. The apparatus consists of a
($1008\times 1024$ pixels) CCD camera, a synchronized stroboscope
illuminating the cell from behind, and specialized image processing
hardware and software from Dantec Instruments. The depth of field of
the camera lens is $\approx 5$ mm. A large cross section of the cell
is imaged  so that several thousand particles can be simultaneously
studied. Velocity maps consisting of $62\times 62$ vectors are
extracted by comparing two closely timed pictures using standard PIV
techniques. Each
 vector is the average velocity of two to four spheres.

Figure \ref{fig:piv} shows typical velocity vector maps  from a
stable fluidized bed, where (a) corresponds to a position near the
top and (b) to a position near the middle of the particle column.
For scale, we also show the magnitude of the Stokes settling
velocity $v_0$. Both velocity maps show regions moving upwards and
downwards, and, as in all of our samples, the magnitudes of the
velocities are significantly larger near the middle than near the
top.

    To quantify these observations,
    we measure the velocity maps  at different
    heights $z$ along the particle column. We
    extract from the velocity maps the mean
velocities,  ${\overline v_z}=\langle v_{i,z}\rangle$, and the root
mean square (rms) velocity fluctuations, $\sigma_v^z=\langle
(v_{i,z}-{\overline v_z})^2\rangle^{1/2}$, where $\langle \ldots
\rangle$ represents an ensemble average over $\sim 50$ vector maps
of $3844$ vectors each.

\begin{figure} \includegraphics[width=80mm]{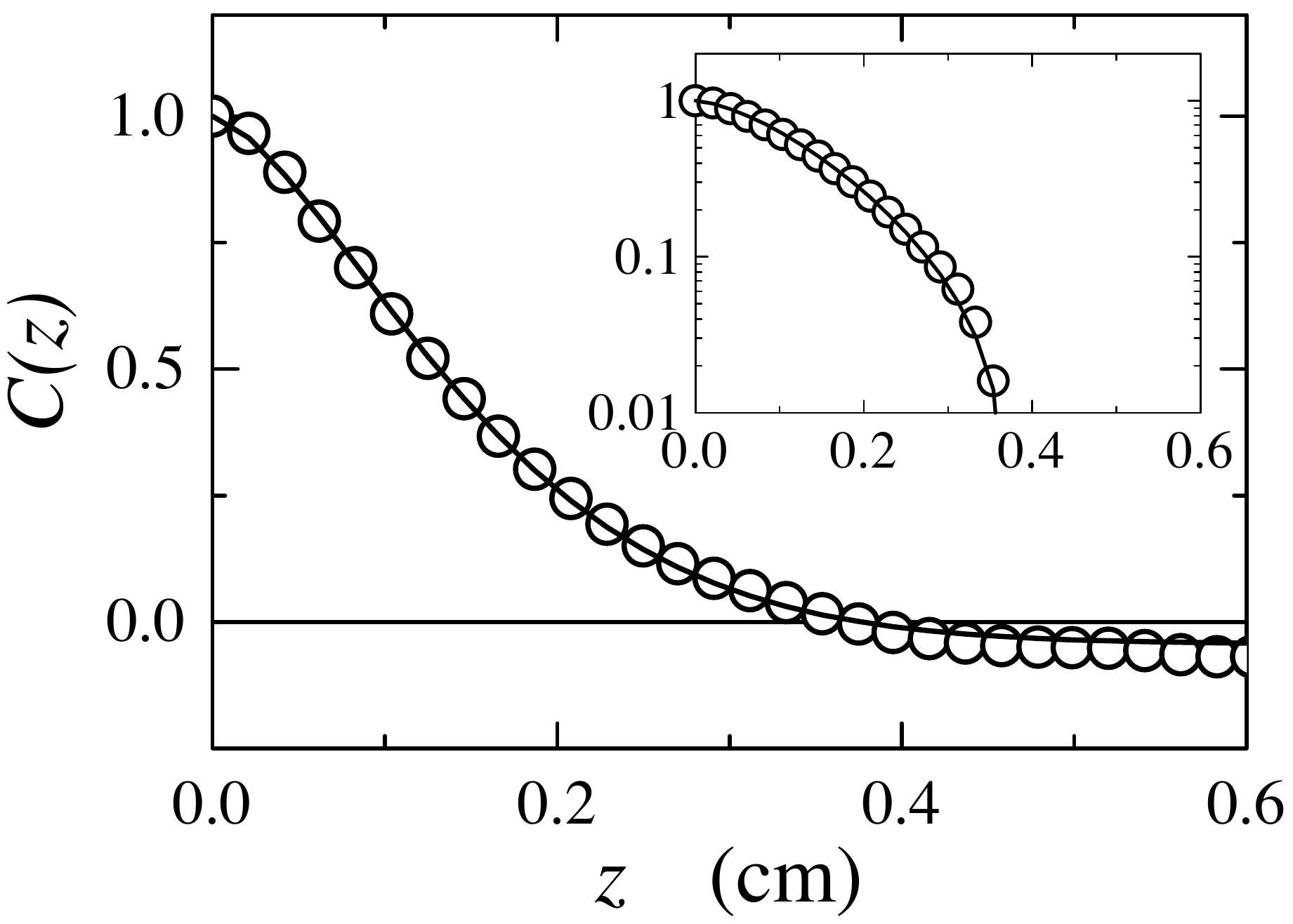}
\caption{Correlation functions of the velocity fluctuations $C(z)$
as a function of distance $z$, near mid-height in the $a=54.5
\;\mu$m particle fluidized bed. The inset shows the logarithm of
$C(z)$, the solid line being a fit to $C(z)=\exp(-(z/\xi)^{1.5})$,
from which we extract values of the correlation lengths $\xi$.}
\label{fig:cf1}
\end{figure}

The typical velocity vector maps in
    Fig. \ref{fig:piv} also  display large regions where the velocity
    vectors are spatially correlated. To quantify this, we
    calculate the normalized spatial correlation functions
    of the vertical velocity $v_z$, defined as
    $C(z)=\langle v_{z}(0)v_{z}(z)\rangle/\langle v_z(0)^2
    \rangle$. Figure \ref{fig:cf1} shows typical results for $C(z)$,
    measured near mid-height in
    a fluidized bed of $a=54.5 \;\mu$m particles.
    The inset shows that the empirical form    $C(z)=\exp{(-(z/\xi)^{1.5})}$
     fits the data well, enabling us to extract values for $\xi$,
    the characteristic (vertical)  correlation length
    of the velocity fluctuations.
\subsubsection{Fluidized Bed Stability}
\begin{figure} \includegraphics[width=85mm]{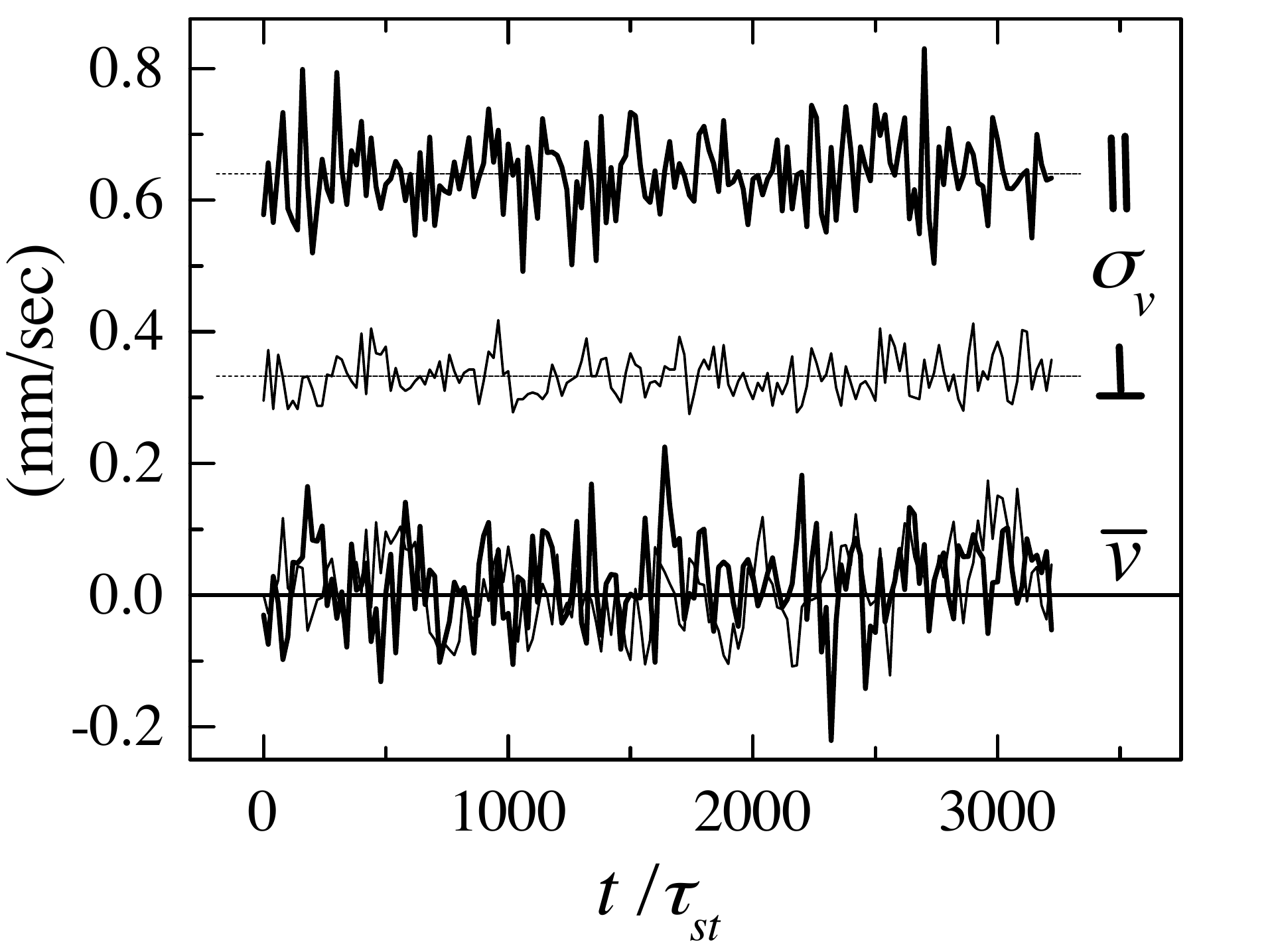}
\caption{Fluidized bed stability: Average velocities ${\overline
v}$, and velocity fluctuations $\sigma_v$, parallel (thick lines) and
perpendicular (thin lines) to gravity as a function of time
$t/\tau_{st}$, where $\tau_{st}=a/v_0$ is the Stokes time. The imaging window is near mid-height in a  fluidized bed of
particles of mean size $a=154.5
\;\mu$m.} \label{fig:stable}
\end{figure}

Before using the fluidized bed apparatus for data collection we need
to ensure that the bed is stable in time and that there are no large
scale convective currents indicative of an unbalanced (i.e. tilted)
cell. To do this we use the PIV imaging method and examine particle
velocities over different regions in the cell. Figure
\ref{fig:stable} shows typical results for ${\overline v}$ and
$\sigma_v$, both perpendicular and parallel to gravity, from PIV
vector maps spanning a $4\times 4$ cm imaging window near
mid-height. All quantities randomly fluctuate, but they also
maintain time-invariant mean values indicative  of a stable system.
Additionally, both velocity averages fluctuate about ${\overline
v}=0$, explicitly demonstrating that particle sedimentation
downwards is
 balanced by the fluidizing counter flow upwards.
 During the experiments, we monitor the mean velocities to ensure
 stability, and we also
    record the position of the top particle/fluid
    interface, clearly visible in Fig. \ref{fig:photobed}, to ensure
    that it does not rise or fall in the cell over time.

\begin{figure} \includegraphics[width=84mm]{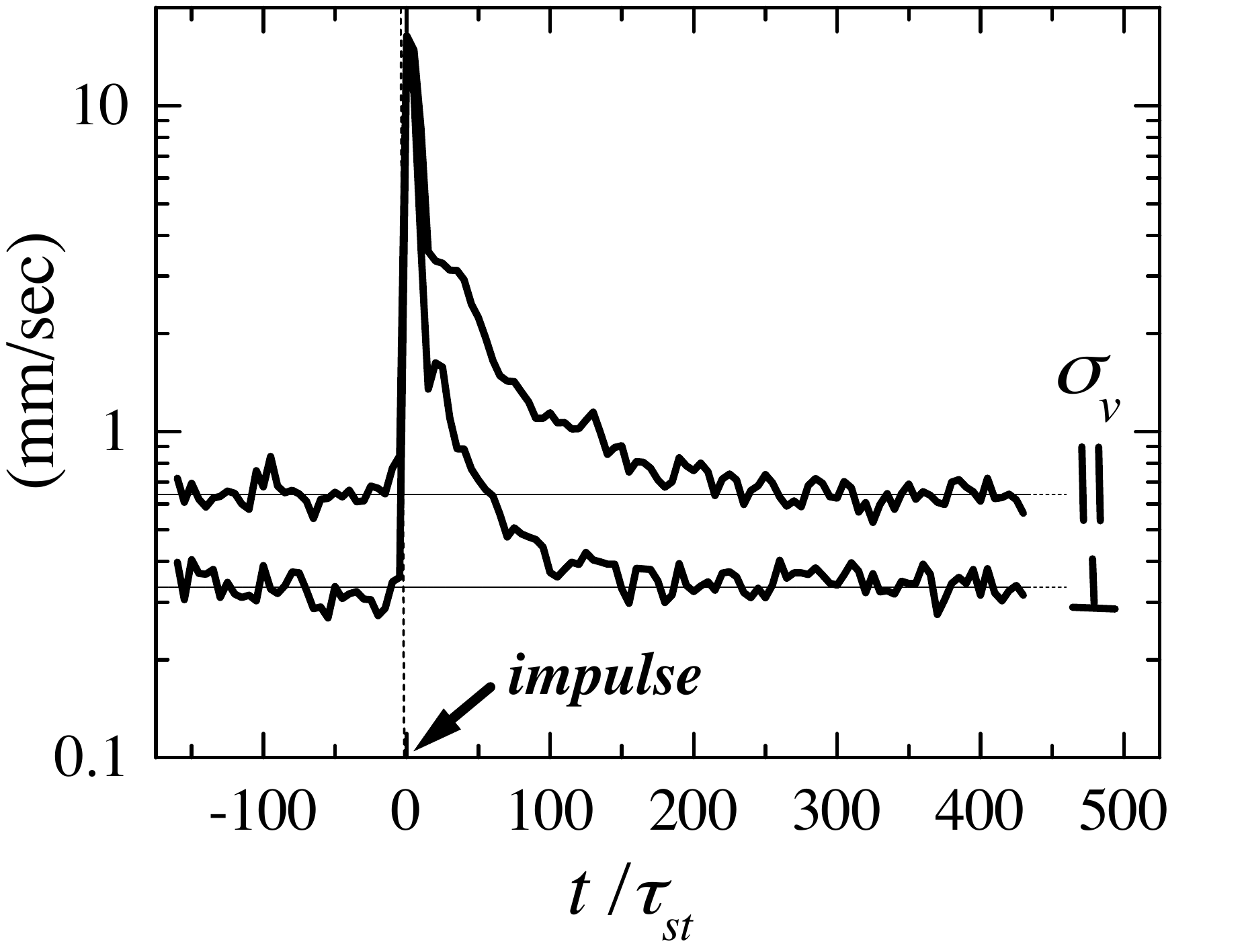}
\caption{Response of a stable fluidized state to a large
perturbation. The velocity fluctuations  $\sigma_v$, both parallel
and perpendicular to gravity, as a function of time $t/\tau_{st}$. At
time $t=0$ a  rapid and large scale impulse is applied throughout
the particle bed to disrupt the particles. }
\label{fig:dvtime}
\end{figure}

As a way of further demonstrating the stability of the bed, we apply
an uncontrolled shock to the system and examine whether or not it
returns back to its original state.   Figure
    \ref{fig:dvtime} shows the test results  in which
     we inject a large ($\sim 3$ cm radius) air bubble
     into the middle of the particle
    bed to cause a rapid and large scale disruption to the particles.
     The particle velocity fluctuations $\sigma_v$, from a $4\times 4$ cm
    subsection of the cell at mid-height, are driven to values at
    $t=0$ that are more than $10$ times their values before the shock.
    Nevertheless, this disturbance rapidly decays away until, for
    times  $t\gtrsim 250 \tau_{st}$, the system returns to its prior
    stable state.
\subsection{Light Scattering System; Particle Concentrations.}
\begin{figure} \includegraphics[width=82mm]{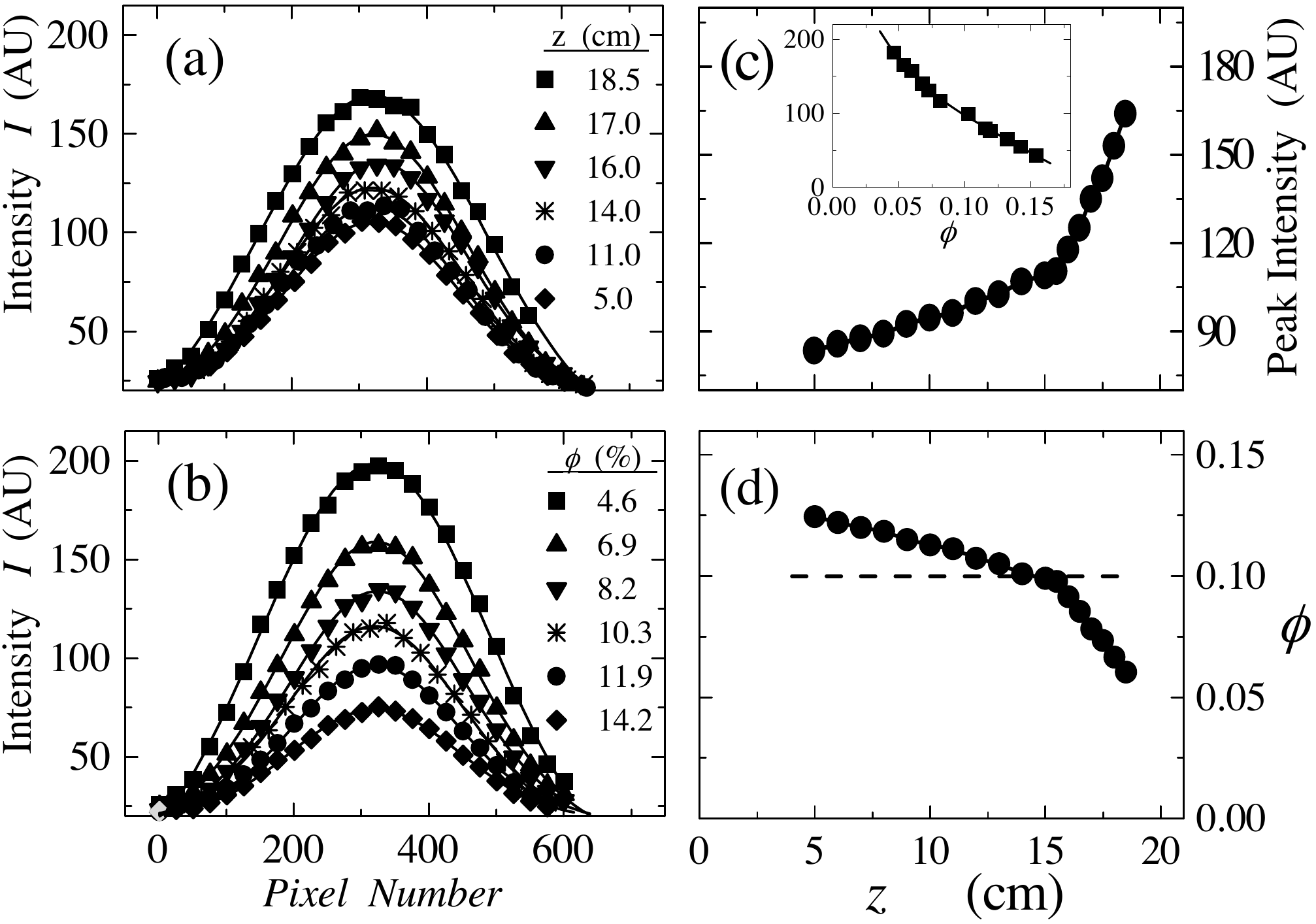}
\caption{(a) Transmitted laser intensity I vs. height $z$ through a
fluidized bed of radii  $a=103.5$ nm particles at average volume
fraction $\phi_0=0.10$.  The solid lines are fits to Gaussian
functions. (b) Intensity vs. $\phi$ calibration. Transmitted laser
intensities measured at mid-height in fluidized beds of average
volume fraction $0.046\le \langle \phi \rangle \le 0.142$ as
labelled. For clarity, only every $25$'th point is plotted. The
solid lines are fits to Gaussian functions. (c) Peak transmitted
intensity $I_{pk}(z)$ vs. height $z$ as extracted from Gaussian fits
to the data in (a). Inset: Calibration curve of peak transmitted
intensity $I_{pk}(\phi)$ vs. $\phi$ as extracted from Gaussian fits
to the data in part (b). (d) Volume fraction $\phi$ vs. height $z$
as determined by comparing the $I(z)$ curve in (c) with the
calibration curve $I(\phi)$ in the inset of (c).}
\label{fig:iandphi}
\end{figure}

    Local particle volume fractions are determined from
     the local optical turbidity
    through the particle column. We pass
    an expanded He-Ne laser beam, of diameter $\sim 0.5$ cm,
    through the fluidized bed at a particular height $z$, and measure
    the transmitted laser intensity $I_T(z)$ using a CCD camera.
    Results for a typical sample are shown in Fig.
    \ref{fig:iandphi}. Figure  \ref{fig:iandphi}(a) shows the measured intensity profiles
    on the CCD at different heights in the particle column. It's evident that
    the transmitted intensity varies with height, with the highest
    intensity at the top, and the lowest at the bottom. To find
    the corresponding particle concentrations, we need a
    calibration reference for the dependence of the transmitted
    intensity on particle concentration. To do this, we make several
    reference fluidized beds of differing average concentration
    $\phi$ and measure the transmitted intensity patterns at
    mid-height. The concentration values are calculated from the
    ratio of the heights of the sedimented particles, at $H_{sed}$ and  $\phi=0.638$,
    to the height of the fluidized column, $H$, as
    $\phi=0.638*(H_{sed}/H)$. The results are shown in Fig.
    \ref{fig:iandphi}(b). No efforts are made to model the
     observed dependence of $I$ on $\phi$, our aim being simply to have a
    calibration reference. Fits of $I(z)$ and $I(\phi)$ in Fig.
    \ref{fig:iandphi}(a) and (b) to Gaussian functions yield peak
    intensity values $I_{pk}$, plotted in
    Fig. \ref{fig:iandphi}(c). By comparing the measured values of
    $I_{pk}(z)$ in our fluidized bed with the reference values of
    $I_{pk}(\phi)$ we are able to determine the height dependent
    concentrations $\phi(z)$ in the fluidized beds, as shown in Fig. \ref{fig:iandphi}(d).
\section{Results}\label{sec:results}
In this section we describe our main results for the particle
dynamics of $4$ fluidized beds that all share the same average
particle concentration, $\phi_0=0.10$, and column height, $H=18.5$
cm. The beds differ, however, in the size of the fluidized
particles, which range from $54.5\le a\le 245\;\mu$m. As a
consequence, the beds differ in two significant ways. First, the
normalized bed heights, $H^*\equiv H/a$, range from  $755$ for the
largest particles to $3364$ for the smallest.  Second, the total
number of particles fluidized, $N=[(3/4\pi)\phi_0 TWH]\times
a^{-3}$, range from $192,000\le N\le 17,500,000$, as listed in Table
II. A comparison of the results for different particle sizes will
allow us to examine the specific dependencies of fluidization on the
system size $N$, and the effective column height $H^*$. For each
fluidized bed we present results for the height dependence of the
particle concentration $\phi(z)$, the velocity fluctuations
$\sigma_v(z)$, and the velocity correlation lengths $\xi(z)$.
\begin{table}
\caption{Measured properties of the fluidized beds described in
Table I. $a$ is the particle radius and $N$ the total number of
particles fluidized. $v_{p}/v_o$ is the normalized fluidizing pump
velocity upwards. $v^{top}/v_o$ is the initial sedimentation
velocity of the top interface, measured just after the fluid pump is
turned off. $\phi^{top}$ is the particle concentration measured just
below the top interface. The particle concentrations
$\phi^{top}_{p}$ and
 $\phi^{top}_{sed}$ are calculated from $v_{p}/v_o$ and
$v^{top}_{sed}/v_o$ using the Richardson-Zaki equation
$v/v_o=(1-\phi)^{5.5}$. }
\begin{tabular}{c|c|c|cc|c|cc}
 $a (\mu m)$ & $H^*$&$N$&$\frac{-v^{top}_{sed}}{v_o}$&$\frac{v_{p}}{v_o} $&$\phi^{top}$&
 $\phi^{top}_{sed}$&$\phi_{p}$\\
\tableline \tableline
$245$&$755$&$1.92\cdot 10^5$&$0.605$&$0.600$&$8.4\;\%$&$8.7\; \%$ &$8.9 \; \%$\\
$154.5$&$1194$&$7.67\cdot 10^5$&$0.659$&$0.654$&$7.0$&$7.3$&$7.4$\\
$103.5 $&$1769$&$2.55\cdot 10^6$&$0.727$&$0.686$&$6.0$&$5.6$&$6.6$\\
$54.5$&$3364$&$1.75\cdot 10^7$&$0.797$&$0.776$&$3.7 $&$4.0\; \%$ &$4.5 $\\
\tableline
\end{tabular}
\end{table}
\subsection{Concentration Profiles $\phi(z)$.}\label{sec:phi}
\begin{figure} %\includegraphics[width=85mm]{prl-segre-fig2.pdf}
\includegraphics[width=89mm]{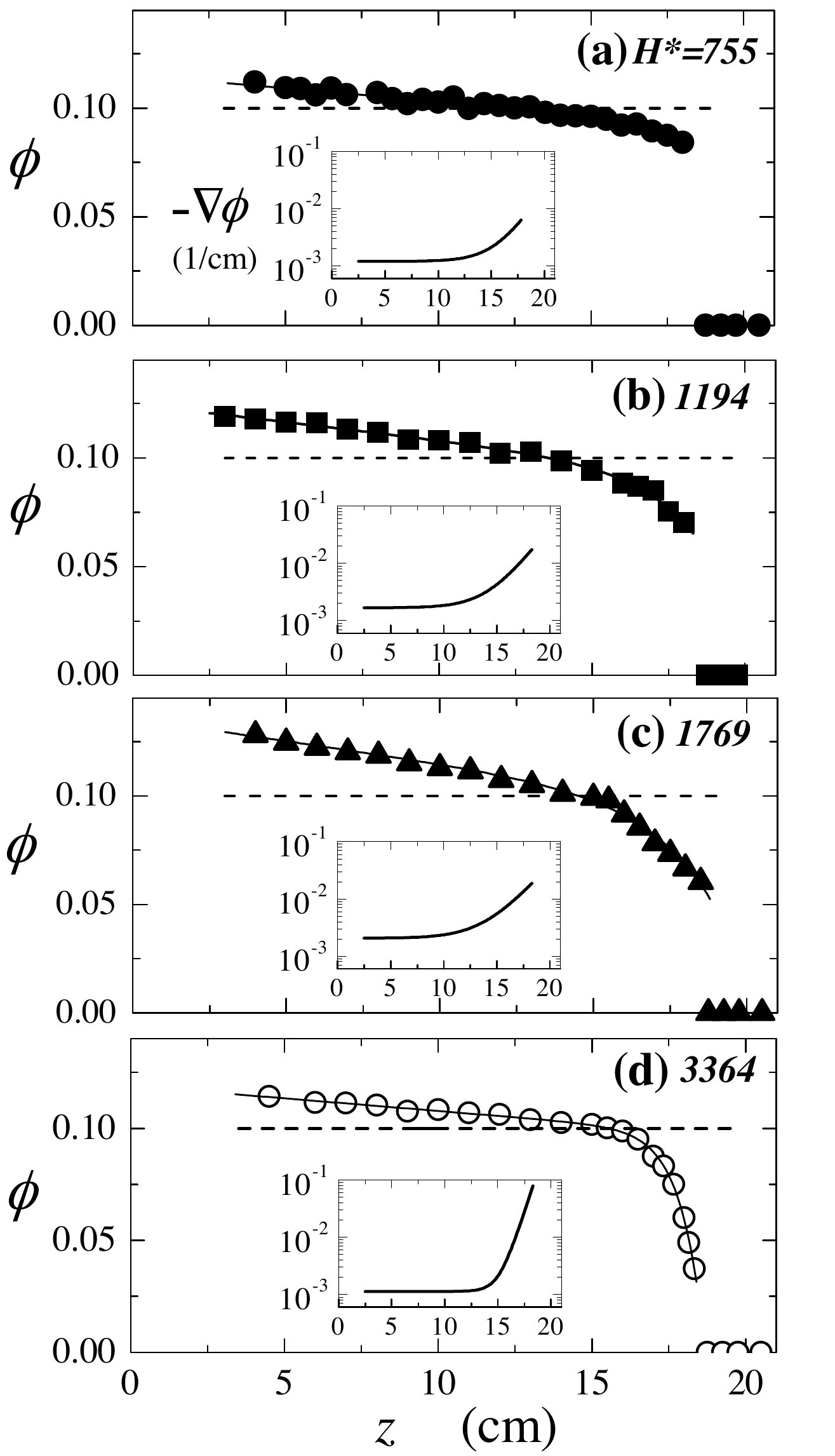}
 \caption{Steady state concentration profiles $\phi(z)$. The particle radii, normalized column height, and number of particles
are (a) $a=245\;  \mu$m, $H^*=755$, $\; N=1.92E5$, (b) $154.5, 1194, 7.67E5$, (c) $103.5, 1769, 2.55E6$, and (d) $54.5, 3364, 1.75E7$.
 The dashed lines show the mean
concentrations, $\phi_0=0.100$. Solid lines are fits to the
empirical form $\phi(z)=b_0-b_1 z-b_2e^{b_3z}$, where $b_0,b_1,b_2$
and $b_3$ are adjustable fit constants.
 Insets: concentration
gradients, $-\nabla \phi =b_1+(b_2b_3)e^{b_3z}$.}\label{fig:4phi}
\end{figure}

We begin by examining the concentration profiles $\phi(z)$ of the
four different particle size fluidized beds. The laser scattering
method, described in section II.D above, allows for the accurate
determination of the local particle
 volume fraction $\phi(z)$ as a function of height $z$. To do this, we first
fluidize each of the four different particle sized samples to the
same mean concentration, $\phi_0=0.100$, and the same total height,
$H=18.5$ cm. After waiting for transients to die out (a time $>
H/v_0\sim 5$ min) we obtain  the time invariant concentration
profiles shown in Figs.  \ref{fig:4phi}(a-d).

It is immediately apparent that none of the fluidized beds are
perfectly homogeneous in concentration. Each concentration profile
shows the same trend for the concentration to decrease with height,
i.e. they are all stratified. While most of the concentration
variation occurs in the highest portions
    of the columns, measurable and significant gradients occur at all heights.

Perhaps the most striking result in Fig. \ref{fig:4phi} is that the
 concentration profiles and the degree of stratification change
with normalized bed height $H^*\equiv H/a$.  That is, despite the
fact that all of the fluidized beds share the same average
concentration, $\phi_0=0.100$, and the same total height $H=18.5$
cm, there is no universal curve for the concentration profiles.
Rather, the profiles are found to  markedly change with normalized
height $H^*$. The main trend in evidence, as described further
below, is for the degree of stratification to increase with
increasing cell height $H^*$, or system size $N$.
\subsubsection{Interfacial Concentrations.}
\begin{figure}\includegraphics[width=87mm]{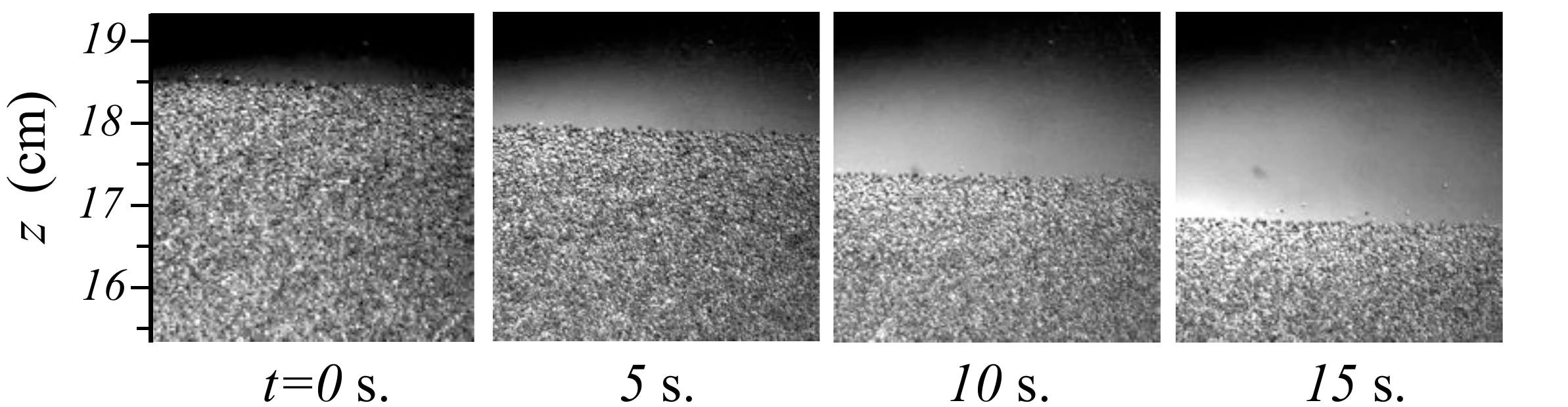}
\caption{Photographs of the top portion of
a particle column, comprised of $a=154.5 \;\mu$m spheres, taken at
different time intervals $t$ {\it after} having turned off the
fluidizing pump.} \label{fig:vsedmeas}
\end{figure}

Before examining the details of the gradients in concentration
$\nabla \phi(z)$ from Fig. \ref{fig:4phi}, we first find it
instructive to consider a  simpler measure of the degree of
non-uniformity in the beds by considering a measure of the variation
of $\phi(z)$ in each column. The smallest values of $\phi(z)$ are
always recorded at the tops of each column, just below the sharp
interfaces (see Fig. \ref{fig:photobed} and \ref{fig:vsedmeas}). To
within our resolution of $\sim 0.5$ cm, there is a near
discontinuity as the concentration drops from a finite value to zero
over $\sim 0.5$ cm. (Note that much more diffuse interfaces occur
for more dilute and/or polydisperse systems \cite{salin,fbprl,front}). We
therefore define $\phi^{top}$ as the value of $\phi(z)$ measured
closest to, and just below, the top interface.
\begin{figure}\includegraphics[width=84mm]{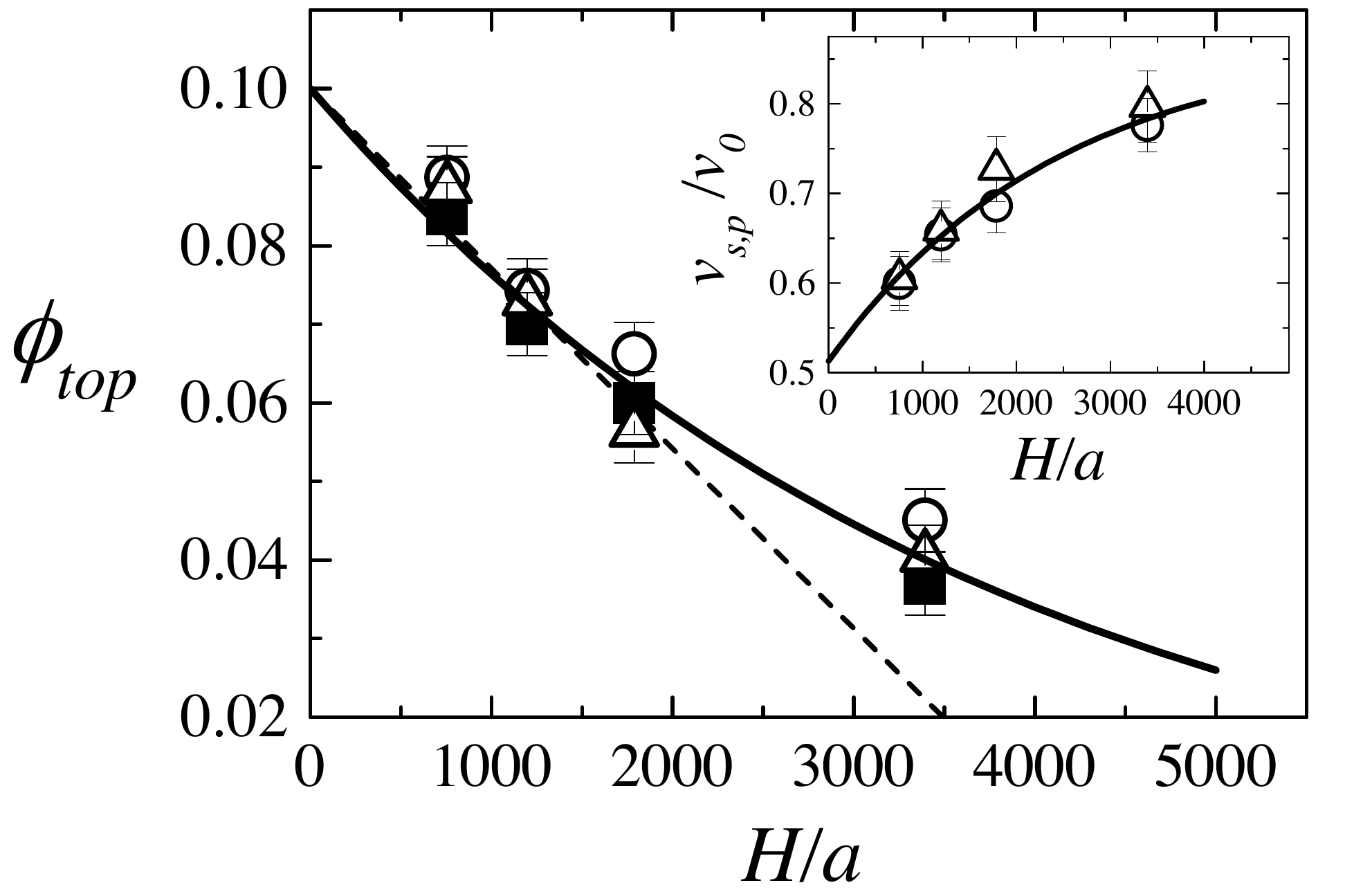}
\caption{The concentration just below the top interface $\phi^{top}$,
 $\circ$, vs. normalized column height $H^*$, from Fig.
\ref{fig:4phi}. Also shown are the  volume fractions $\phi$
calculated  from measurements of the settling velocities of the top
interface, $v_{sed}^{top}$, $\blacksquare$, and the fluidization
pump velocity $v_p/v_0$, $\bigtriangleup$, using the RZ equation
$v(\phi)/v_0=-(1-\phi)^{5.5}$. Inset: The initial settling velocity
of the top interface, $-v_{sed}^{top}$, $\blacksquare$, and the
fluidization pump velocity $v_p/v_0$, $\bigtriangleup$, as a
function of column height $H^*$.}\label{fig:vsedtop}
\end{figure}

Figure \ref{fig:vsedtop} shows our measured values of $\phi^{top}$
vs. the normalized bed height $H^*$. The overall trend is clear, the
taller the column $H^*$,  the smaller the concentration
$\phi^{top}$, and the greater the degree of stratification. For our
$3$  shortest columns, there is a near linear decrease of
$\phi^{top}$ with $H^*$. This linear trend, of course, cannot
continue indefinitely, as $\phi^{top}$ cannot go below zero. Indeed,
in our tallest bed, the result  deviates significantly from the
initial linear trend.

    To help corroborate the concentration values $\phi^{top}$
just below the top interface, we also conduct a second series of
measurements. Because the hindered settling function of a collection
of spheres is strongly $\phi$-dependent (the Richardson-Zaki (RZ)
equation \cite{rz} is $v(\phi)/v_0=(1-\phi)^{5.5}$), by measuring
the settling velocity of the top interface, $v_{sed}^{top}/v_0$, we
can independently extract out  values for $\phi^{top}$.
    To test this, we first establish a
    stable fluidized bed, then turn off  the fluidizing pump
    and measure the initial settling rate of the top interface,
    $-v_{sed}^{top}/v_0$. Figure \ref{fig:vsedmeas} illustrates how these
    experiments are done. Results for $-v_{sed}^{top}/v_0$ vs.
    $H^*$ are shown in the inset of Fig. \ref{fig:vsedtop}. The
    corresponding (RZ calculated) values of $\phi^{top}$ are shown
    in the main portion of the Figure, and agree to within the error
    bars with the turbidity measurements.
     In this way, we have a second, independent, way of
     determining $\phi^{top}$ that further corroborates the non-universality of the
     concentration profiles with particle size $a$.
\subsection{Concentration Gradients $\nabla \phi(z)$.}\label{sec:gphi}
\begin{figure} \includegraphics[width=80mm]{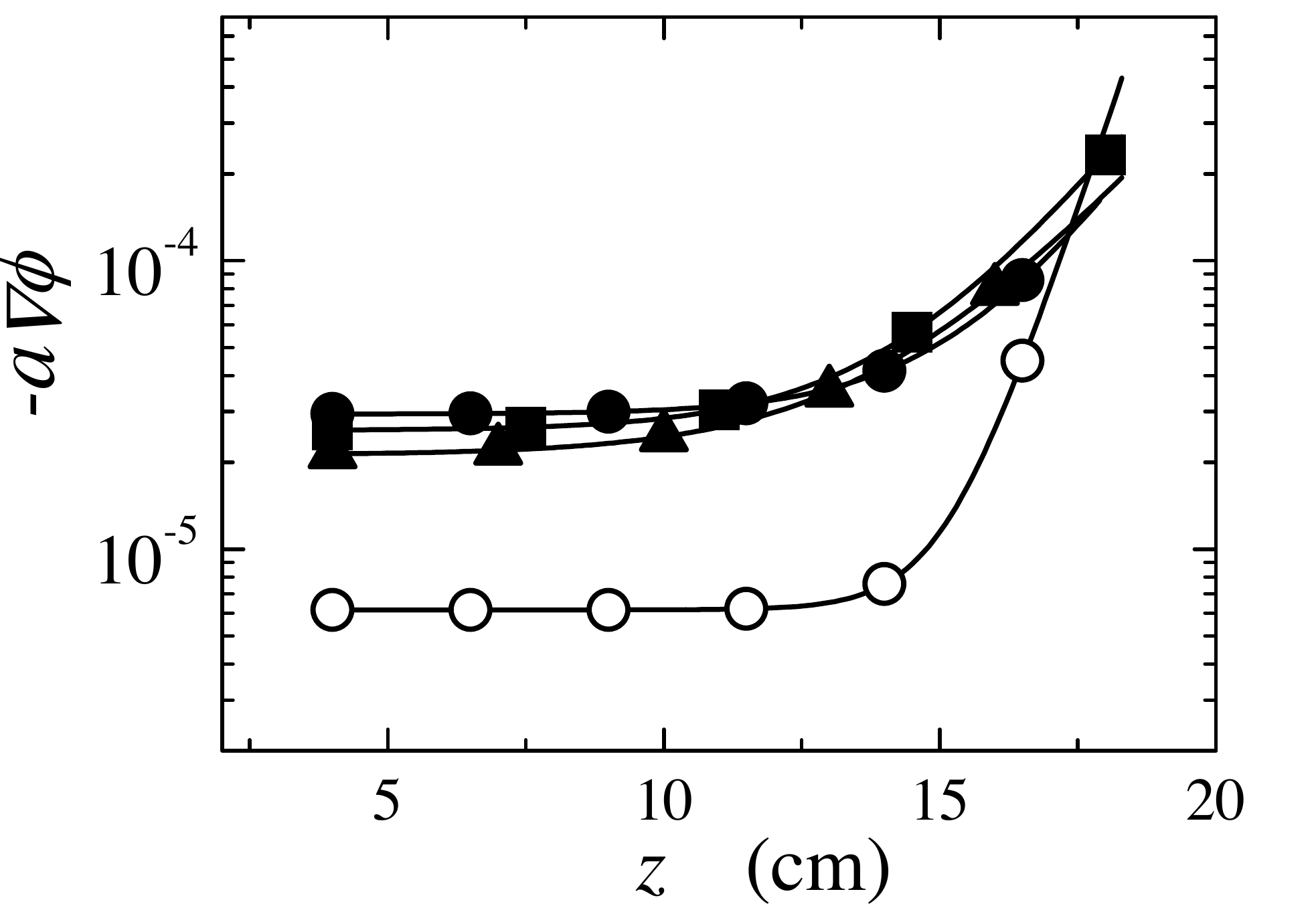}
\caption{Scaling of the concentration gradients, $-a\nabla\phi$ vs.
height $z$, from fits to the $\phi(z)$ profiles shown in Fig.
\ref{fig:4phi} ($-a\nabla \phi =ab_1+(ab_2b_3)e^{b_3z}$). The
smallest particle size deviates from scaling because the gradient
cannot exceed its maximum value (see text). The different symbol
shapes correspond to differing particle sizes as defined in Fig.
\ref{fig:4phi}.}
 \label{fig:mids}
\end{figure}

 The insets in Fig. \ref{fig:4phi}(a-d) show the
concentration gradients $-\nabla \phi$, as derived from fits to
$\phi(z)$. The gradients are all nearly constant up to  mid-height,
and increase markedly thereafter. They also increase in magnitude
with decreasing particle size $a$.  To examine this in closer detail,
we plot in Fig. \ref{fig:mids} the scaled
 values $-a\nabla\phi$ vs. height $z$. Remarkably, data for
 the $3$ largest particle sizes collapse closely together,
 showing that in this regime there is a new scaling relation, $-\nabla \phi\propto 1/a$.
This scaling also shows that the stratification
increases with system size. Since the total number of particles in our samples, $N\propto 1/a^3$, we can equivalently recast the scaling form as $-\nabla \phi\propto N^{1/3}$. Significantly, this form predicts that in limit of small system sizes, $N\rightarrow 0$, the stratification vanishes so that the sample becomes homogeneous in concentration.

It is also readily apparent that the scaling form $-\nabla
\phi\propto 1/a$ predicts an un-physical divergence of $-\nabla
\phi$ in the small particle limit $a\rightarrow 0$. Consequently,
the scaling form $-a\nabla \phi$ must eventually break down for
small enough particles, and  indeed, we do see large a large
deviation from scaling in Fig. \ref{fig:mids} for our smallest
particle size, $a=54.5 \;\mu$m. To help explain this, we note that
the largest (linearly averaged over all heights $z$) value of
$-\nabla \phi$ that is physically possible, for a bed of height $H$
and average concentration $\phi_0$, is $\left<-\nabla
\phi\right>_{max}\sim 2\phi_0/H$. Our height averaged gradients in
the scaling regime, $-\left<\nabla \phi(z) \right>\equiv
\frac{1}{H}\int_0^H\nabla \phi(z)dz \sim 4.6\cdot 10^{-5}/a$, will
reach this maximum gradient when the particles become small enough,
of order $a^*\sim 4.6\cdot 10^{-5}H/2\phi_0\sim 43 \; \mu$m. By
these arguments, the $-a\nabla \phi$ data collapse cannot be
maintained for bead sizes $a\lesssim a^*=43 \;\mu$m, because this
would require $\left<-\nabla \phi\right>$ to exceed its maximum
value $2\phi_0/H$. Consequently, for beads close to or smaller than
$a^*$, plots of the product $-a\nabla \phi$ should deviate below
those seen in the scaling regime, exactly as we see in Fig.
\ref{fig:mids} for our only set of beads that are close in size to
$a^*$.
\subsubsection{Stratification Prediction Resulting from Particle Size
Polydispersity} Before continuing further, we  critically examine
the degree to which our results for $\nabla \phi(z)$ may be
influenced by the non-zero particle size polydispersity $\sigma_a/a$
of our beads. In general,  a concentration stratification can arise
due to a combination of (1) an inherent physical process found in
monodisperse beads, and (2) purely the effects of particle size
polydispersity. Several relevant points can be made for our systems.
Firstly, we note that all $4$ particle sizes used  have similar
levels of polydispersity, with $1.1\le \sigma_a/a\le 1.6 \%$, yet
the degree of stratification varies greatly between the samples,
suggesting that polydispersity is not the root cause. Additionally,
recent computer simulations \cite{tonypoly} compared the
sedimentation dynamics and structure of particles with
polydispersities ranging from $0\%$ (perfectly monodisperse) up to
$10\%$. While differences between the monodisperse and $10\%$ cases
were found, a polydispersity of $2\%$, slightly larger than our
samples, yielded results that were, to within errors, identical to
the monodisperse case.   These arguments suggest that particle size
polydispersity is not the dominant cause of the observed
concentration gradients.

\subsection{Velocity Fluctuations $\sigma_v$. }
\begin{figure}\includegraphics[width=84mm]{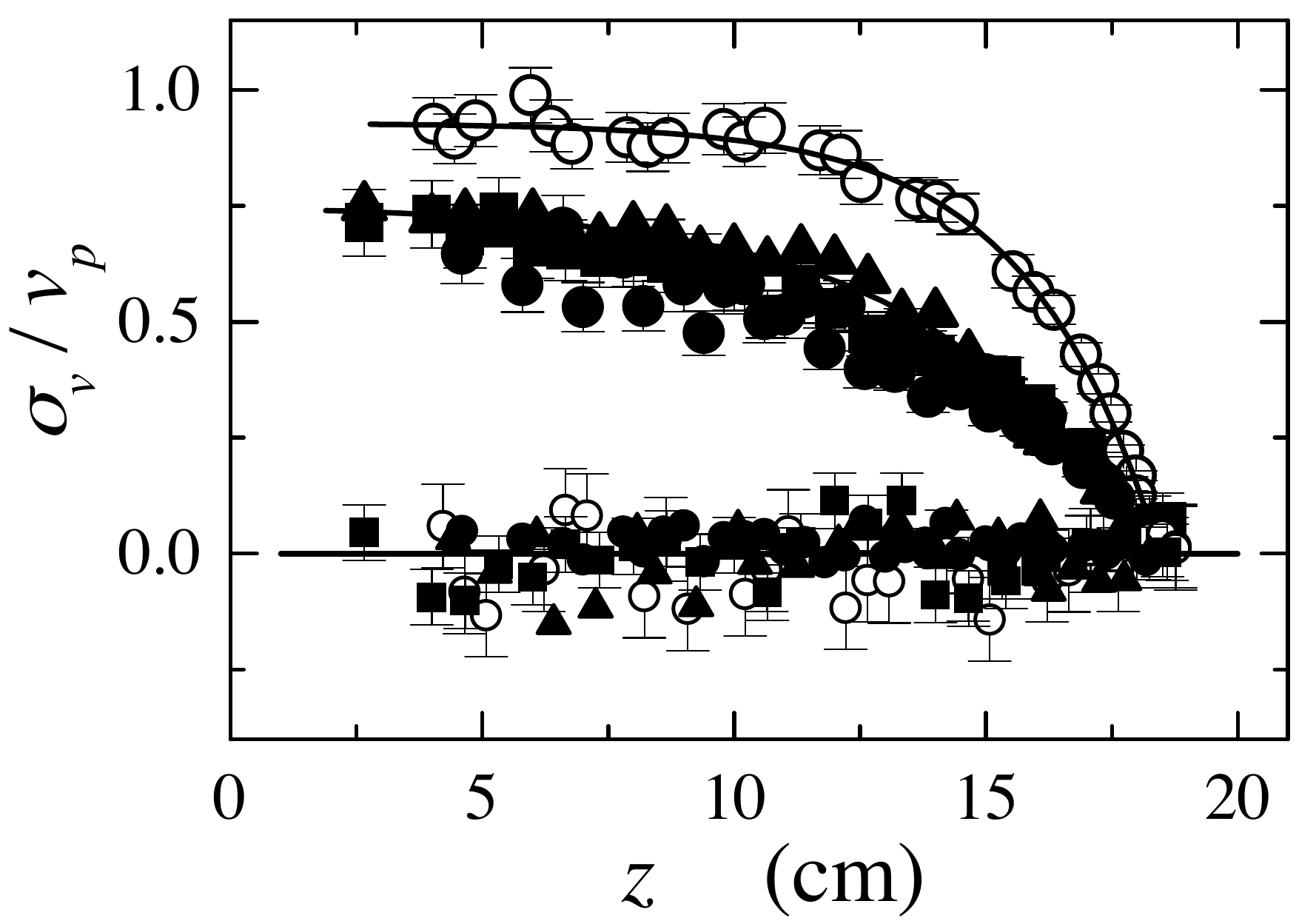}
 \caption{Normalized
velocity fluctuations $\sigma_v$, large symbols, and the mean
particle velocity ${\overline v}$, small symbols, parallel to gravity as a function of height $z$ in stable fluidized beds. The different symbols
correspond to the different particle sizes as defined in Fig.
\ref{fig:4phi}. The solid lines are guides to the
eye.}\label{fig:dv}
\end{figure}

To measure the particle velocity fields, we use
    PIV methods. Figure \ref{fig:dv} displays the values
    of the particle velocities for the $4$ different particle size fluidized beds.
Measurements are taken at multiple heights from near to the bottom
up to the top, at a height of $H\sim 18.5$ cm. In all $4$ samples the
average velocities ${\overline v}\sim 0$, indicative of stable
fluidization. Results are also shown for the normalized velocity
fluctuations, $\sigma_v/v_p$, for the four different particle size beds.

    In all samples, the fluctuation amplitudes decrease
    with height $z$ in the column. At the top, fluctuations
    become vanishingly small, i.e. the particles are barely moving,
    which we also
confirm by direct observation.

The behavior of the velocity fluctuations also mirrors that of the
concentration gradients. That is, as found for $-a\nabla \phi$ in
Fig. \ref{fig:mids}, Fig. \ref{fig:dv} shows (i) a scaling regime, for $a\gg a^*$, in which the fluctuation values  $\sigma_v(z)/v_p$ collapse
together, and (ii) deviations from scaling  for $a\sim a^*$.
\subsection{Velocity Fields-Correlation Lengths}
\begin{figure} \includegraphics[width=80mm]{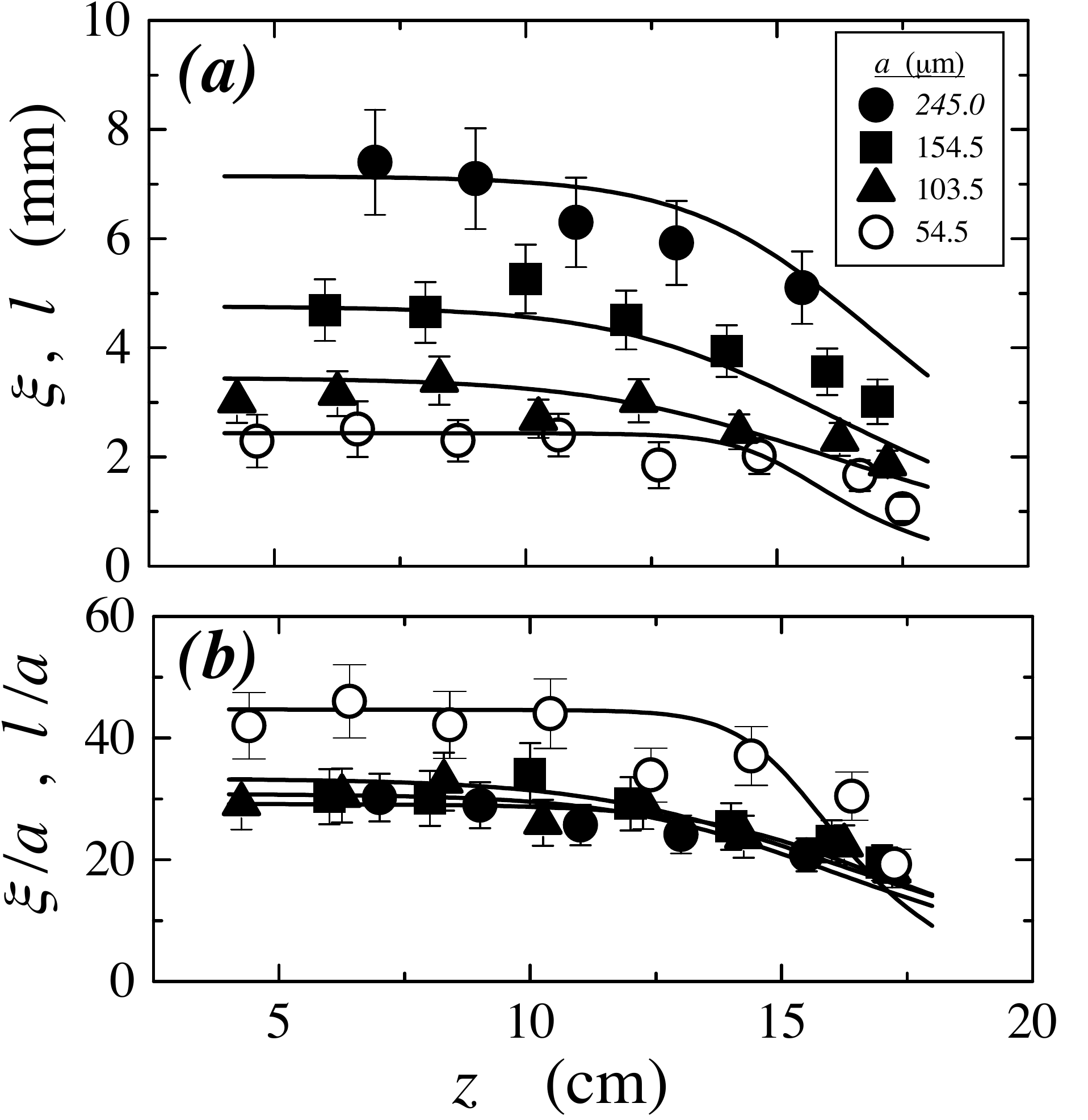}
 \caption{ Velocity correlation lengths plotted as $\xi$ vs. height $z$ in
 (a), and $\xi/a$ in (b). The solid lines in (a) are the
 predictions of MB model for $l$, Eq. (1),  with  fit constants, in order of decreasing particle size, $c_0=0.82, 0.82, 0.82$ and  $0.70$. (b) The MB predictions for $l/a$.}\label{fig:xi}
\end{figure}

We next describe the characteristic lengthscale $\xi$ of the
velocity flucutuations. From the velocity maps as measured by PIV,
we calculate the spatial correlation functions \cite{segreprl} of
the vertical velocity fluctuations,
 $C(z)=\langle v_{z}(0)v_{z}(z)\rangle/\langle v_z(0)^2 \rangle$, fit them to the form $C(z)=\exp{[(-z/\xi)^{1.5}]}$, and extract values for for $\xi$ for the $4$ different particle size fluidized beds are shown in Fig. \ref{fig:xi}({a}).

The correlation    lengths are not uniform in height, and exhibit a
slight    decrease towards the top part of the column \cite{fbprl}.
    Moreover,  at every height $z$, $\xi$
    increases in magnitude with increasing particle size. To
    examine this further, we plot in Fig.
     \ref{fig:xi}(b)  the values of
    the correlation lengths normalized by their particle radii, i.e.
          $\xi/a$.
Significantly, the  correlation lengths show the same behavior as
the concentration gradients and  fluctuation magnitudes.
That is, (i) a scaling regime, for $a\gg a^*$, in which the
correlation lengths $\xi(z)/a$ collapse together, and (ii)
deviations from scaling for $a\sim a^*$.
\subsection{Connections between Fluidization and Sedimentation}
    While in this article we focus entirely on
     the dynamical properties of fluidized particle beds, it is also
     of interest  to understand the similarities between
      fluidization and sedimentation.
      In the lab frame, fluidization
      and sedimentation differ in that fluid is pumped upwards in
      the former, and not pumped at all in the latter. In the particle frame of reference, however, the particles in both cases are moving through a background fluid at a rate determined by their buoyant mass and
      concentration. It is not evident,  {\it a priori}, how the
      particle dynamics differ between fluidization and sedimentation.
\begin{figure} \includegraphics[width=87mm]{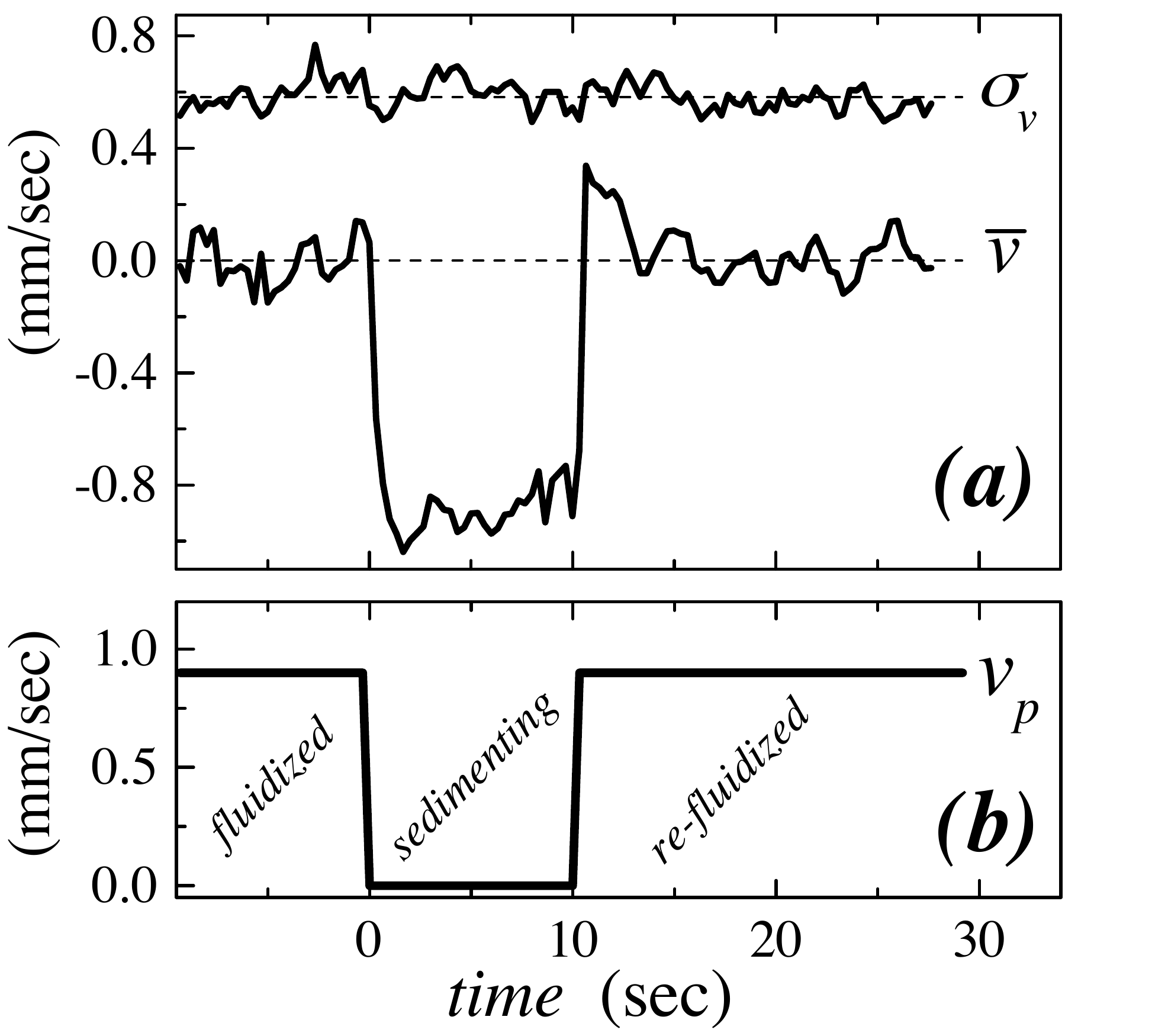}
\caption{Comparison of fluidization and sedimentation. (a) the
velocity fluctuations $\sigma_v$ and mean velocity ${\overline v}$
as a function of time $t$. (b) The upward fluid velocity $v_p$ of
the fluidized bed pump. For times $t<0$, the pump is on, and the
particles are stably fluidized. At time $t=0$, the fluidization pump
is turned off, and the particle sediment downwards until, at $t=10$
s, the pump is turned back on to re-fluidize the system. The
fluctuation magnitudes $\sigma_v$  show no discernible difference
between fluidization and sedimentation.} \label{fig:pumponoff}
\end{figure}

      For a direct comparison on a typical sample, we show in Fig.
    \ref{fig:pumponoff} results for particle velocity fluctuations
    for the same sample in both fluidized and sedimenting states.
    Fluidization occurs for times leading up to $t=0$ sec, when the fluid pump is on, and the average particle velocities in the vertical direction ${\overline v}$ are
very close to zero.  Sedimentation begins at $t=0$ sec., when the
fluid pump is temporarily turned off, and the particles move
downwards with an average velocity $v_z\sim -0.9$ mm/s. (The rapid
response of the particles to the turning on and off of the fluid
pump reflects the lack of particle inertia.) Finally, at $t=10$ s,
after the particles have fallen $\sim 1$ cm (much less than the
initial bed height $H\sim 18$ cm), the fluidizing pump is turned
back on, and the particles are re-fluidized.

The  remarkable feature seen in Fig. \ref{fig:pumponoff} is that the
fluctuation magnitudes $\sigma_v$ show no discernible difference
throughout the transition from fluidization to sedimentation and
back to fluidization. We interpret this as  evidence that the
particle fluctuation dynamics during  fluidization and sedimentation
are very similar. To examine this  further, in the following two
subsections we compare our fluidization results for $\xi$ and
$\sigma_v$ to published results for the same quantities measured in
sedimentation experiments.
\subsubsection{Comparison to $\xi$ from Sedimentation Experiments}
\begin{figure} \includegraphics[width=80mm]{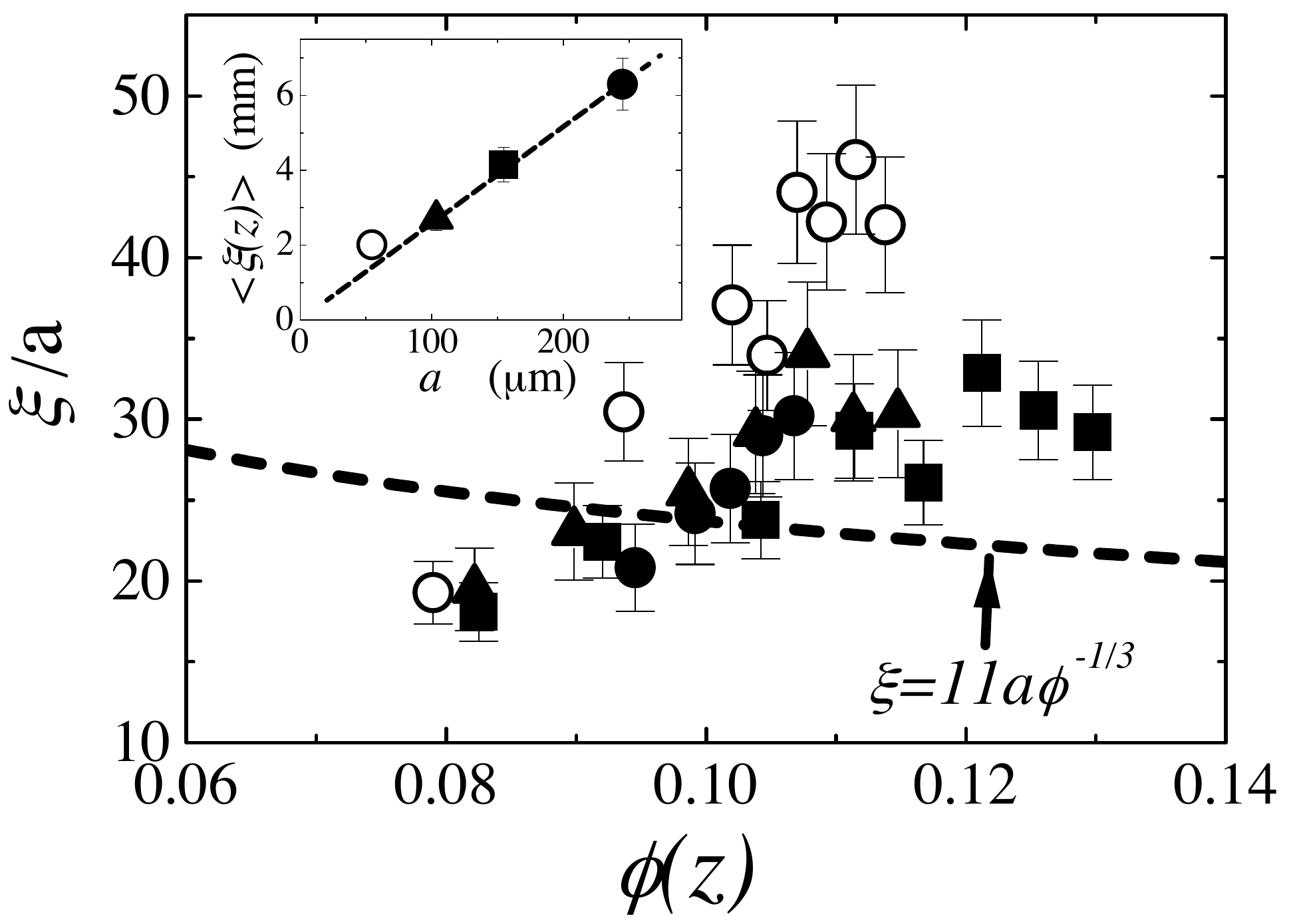}
 \caption{Inset: Height averaged values ${\overline \xi(z)}$
vs. particle size $a$. The dashed line is a fit to ${\overline
\xi}=12a\phi_0^{-1/3}$. Main figure: velocity correlation lengths
$\xi/a$ vs. volume fraction $\phi$.
 The dashed line represents the sedimentation scaling equation, ${\overline
\xi}=11a\phi_0^{-1/3}$, for the height averaged correlation lengths
\cite{segreprl}.}\label{fig:xised}
\end{figure}

Results for the correlation lengths $\xi/a$ in sedimenting samples,
covering a wide range of volume fractions $0.0001\le\phi\le 0.50$,
have been published by Segr{\` e} et {\it al}.
\cite{segreprl,nature}. The methodology by which $\xi$ was measured
differs  from that used here during fluidization. In the
sedimentation experiments, because of the limited time during which
the particle column is falling, it was not possible to measure the
fluctuations at different heights. Rather, a fixed camera recorded
data continuously while the column fell past. This produced a single
value of $\xi$ that represents correlation values {\it averaged}
over a large portion of the column, $0.25\lesssim z/H \lesssim
0.75$.  When results from samples of various average concentrations
$\phi_0$ were collected, the data were consistent with the scaling
relation $\xi_S=11a\phi_0^{-1/3}$.\cite{segreprl,nature} For a
meaningful comparison to our fluidization results, we thus need to
first perform a similar height-averaging procedure to our data from
Fig. \ref{fig:xi},
 i.e. $\left <\xi(z)\right>\simeq\frac{2}{H}\int_{0.25H}^{0.75H}\xi(z)dz$,
 with results plotted in the form $\left<\xi(z)\right>$ vs. $a$ in the inset
  of Fig. \ref{fig:xised}. The data fit well to $\left<\xi(z)\right>\approx (25.8\pm 2)a$,
  in remarkably good agreement with the scaling relation from sedimentation,
  $\xi_S=11a\phi_0^{-1/3}=23.7a$.

 We emphasize again that the scaling relation found in sedimentation,
 $\xi_S=11a\phi_0^{-1/3}$, represented a single, height-averaged,
 correlation length for a column of mean concentration $\phi_0$.
 By comparison, our fluidization results  in Figs. \ref{fig:4phi}
  and \ref{fig:xi} display the complete
  height dependence of both $\xi$ and $\phi$ from {\it within} a single column.
By combining these data, we can ask a new question. Can  the scaling
 form $\xi=11a\phi^{-1/3}$  be used to predict the local values of
 $\xi$ from the local values of $\phi$ {\it within} a single column?
 To answer this, we plot in Fig. \ref{fig:xised}  results from all
 four fluidized beds for $\xi(z)/a$ vs. the corresponding local
 volume fractions $\phi(z)$. It is immediately apparent that the scaling form
 does not describe the data, and has the opposite dependence on
 $\phi$. Moreover the data points themselves do not collapse
 together, suggesting that, unlike the column averaged values
  $\left<\xi(z)\right>$, the values of $\xi$ within a column are not a unique
  function of $\phi$ alone.
\subsubsection{Comparison to $\sigma_v$ from Sedimentation Experiments.}
Results for the velocity fluctuations $\sigma_v$ in semi-dilute sedimenting
samples have been published by numerous groups. Guazzelli
\cite{nandg1} was the first, and found for a sample at the same
concentration used here, $\phi_0=0.10$, that
$\sigma_v/v_{sed}=0.97\pm 0.19$. Later, Segr{\` e} \cite{nature}
measured $\sigma_v/v_{sed}=0.70\pm 0.08$ at $\phi=0.05$, and
$\sigma_v/v_{sed}=0.9\pm 0.1$ at $\phi=0.13$. All of these values
reflect height-averaged values of the falling columns, as described
above. Ladd et al. \cite{tony1} also conducted Lattice Boltzmann
simulations of up to $32,000$ particles sedimenting in a fully
bounded box at $\phi_0=0.13$. They found $\sigma_v/v_{sed}\simeq
0.6$ for the largest systems. All of these sedimentation values are
 in reasonably good agreement with our fluidization results
shown in Fig. \ref{fig:dv}, where near to the middle of the columns,
$0.6\lesssim \sigma_v/v_p \lesssim 0.9$.

\section{Evaluation of the Mucha and Brenner Stratification Model}\label{sec:sa}
The main aim of  section \ref{sec:results} is to describe results
for the particle dynamics during stable fluidization of a variety of
different sized particle beds. We are not aware of other published
results or predictions, either experimental or theoretical,  for the
specific forms of the quantities we measured, $\phi(z)$, $\nabla
\phi(z)$, $\sigma_v(z)$ or $\xi(z)$, in similar fluidized beds.
Published results are available  for some of these properties in
similar systems that are not fluidized, but rather are sedimenting
downwards. Significantly, we were able to make a connection between
fluidization and sedimentation by showing in Fig.
\ref{fig:pumponoff} that the local velocity fluctuations in a
typical sample are very similar in either mode. A subsequent
comparison  confirms this connection by showing that our average
correlation lengths $\left<\xi(z)\right>$ and velocity fluctuations
$\sigma_v$ in fluidized beds agree quite well with published results
in sedimenting systems. Building upon this link, and in the absence
of relevant fluidization theories, we therefore find it worthwhile
to turn to theories of particle sedimentation to help explain our
data.

    The Mucha and Brenner model \cite{muchaPOF,jfm}, as outlined in the introduction, is
    a recent theory of sedimentation that argues that the
    value of the local velocity correlation lengths $\xi$ are primarily
    determined by the local degree of stratification $\nabla \phi$,
    see Eq. (\ref{eq:bm}). Results in support of this model, from numerical
simulations \cite{muchaPOF}, have been confined to  the transition
region near the top particle interface, but whether it can provide a
complete description of  $\xi$ over the entire sedimenting column
remains an open question. The importance of the MB model, if
verified, is that it would provide a  new insight into the origin of
the heretofore poorly understood fundamental lengthscale $\xi$ that
is ubiquitously seen in both fluidized and sedimenting systems.

      For a complete test of  the MB model,  in each of the four beds
      we       input fits to our data for
      $\phi(z)$ and $\nabla \phi(z)$ into Eq. (\ref{eq:bm}) to
      generate  model predictions for the velocity correlation
      lengths $l(z)/a$. These values are then directly compared to our
      correlation measurements $\xi(z)/a$. Figure \ref{fig:xi} shows
      the results. The solid lines are the MB model predictions, and the symbols
      our data. Significantly, there is very good
agreement between theory and experiment for all four fluidized beds.
The adjustable fit constant $c_0$, which is not specified by the
theory but is expected to be of order $1$, has a single value,
 $c_0=0.82$, for the $3$ scaling samples, $a\gg a^*$. For the
 non-scaling sample, with $a\sim a^*$, the model similarly works well,
 but requires a slightly lower fit value, $c_0=0.70$. The agreement is all the more impressive given that the gradients
$\nabla \phi(z)$ are highly non-linear, strongly varying, functions
of height $z$. The model
  Eq. (\ref{eq:bm}) can also be simplified further by noting that the term
  $[\phi S(\phi)]^{1/5}$ is nearly constant over a wide range of volume fractions,
  $[\phi S(\phi)]^{1/5}=0.52 \pm 0.03$ for $0.04\le \phi\le
  0.30$. In the scaling regime, $a\gg a^*$, therefore
\begin{equation}
\xi(z)/a \approx 0.43[-a\nabla\phi(z)]^{-2/5}.
\end{equation}

 The truly remarkable feature about the MB model is
that it accounts for the characteristic length scale of the velocity
fluctuations, $\xi$, from which the magnitudes of the velocity
fluctuations and diffusion coefficients can be calculated
\cite{segreprl,nature}, {\it solely} from measurements of the
stationary concentration profiles $\phi(z)$. Its validity
corroborates the basic assumption of the model, that the velocity
fluctuations are controlled by the concentration stratification
$\nabla \phi$.
\subsection{Re-derivation of the MB Model}\label{sec:advec}
Finally, we show that the MB model Eq. (\ref{eq:bm}), obtained by
Mucha and Brenner on the basis of physical arguments concerning
buoyancy matching, can also be derived from a general
advection-diffusion model of steady state sedimentation. In this
model, as originally proposed by Kynch \cite{kynch}, the local
particle concentration varies in time as
  \begin{equation} \partial \phi/\partial t=\nabla [\phi
v+D\nabla \phi].  \label{eq:ad1}\end{equation} In our experiments,
we achieve stable fluidization  so that, when averaged over the
short fluctuation lifetimes $\tau$ ($\sim\xi/\sigma_v\sim 5-10$ s),
the concentration profiles $\phi(z)$ shown in Fig. \ref{fig:4phi}
are {\it time-independent} quantities. Expressed mathematically,
$\partial\langle \phi\rangle_\tau/\partial t=0$, where
$\left<\cdots\right>_\tau$ represents an average over times greater
than $\tau$. For a stably fluidized system, Eq. (\ref{eq:ad1})
predicts that the time averaged advective, $J_{A}=\left<\phi
v\right>_\tau$, and diffusive, $J_{D}=-\left<D\nabla \phi
\right>_\tau$,  terms are equal at all positions in the particle
column.

To evaluate the advective term, we  expand to first order,
$\phi(z,t)=\phi(z)+\delta\phi(z,t)$, and
 $v(z,t)=\langle v(z,t)\rangle_\tau+\delta v(z,t)$. (For
 compactness, we will simplify the notation below by not writing the
 explicit $z$ dependence, i.e. $\phi(z)\rightarrow \phi$,
 but all quantities in general depend upon $z$).
 Since the mean particle velocity in a fluidized bed
 vanishes, $\langle v\rangle_\tau=0$, and
 the time averaged fluctuations are zero,
 $\langle \delta \phi\rangle_\tau=0$ and $\langle \delta
v\rangle_\tau=0$, the advective term reduces to $J_{A}=
\left<\delta\phi\delta v \right>_\tau$, which is non zero because
velocity and concentration fluctuations are anti-correlated (denser
regions fall, lighter regions rise) \cite{hinch,bruce}.
If we approximate the  fluctuation terms by their {\it rms} values,
 i.e.  $\delta \phi\sim \sigma_\phi= \sqrt{\left<(\delta \phi)^2
\right>_\tau}$, then
   \begin{equation} J_{A}\approx -\sigma_\phi \sigma_v.
  \end{equation}

To evaluate the diffusive term, $J_D$, we use the published result
(from sedimentation experiments\cite{nature,nandg1}) that the
hydrodynamically induced particle diffusion coefficient is related
to the fluctuations through $D\approx \sigma_v\xi$, so that
  \begin{equation} J_D\approx-\sigma_v\xi\nabla\phi.
    \end{equation}
     For the stability criteria,
$J_A=J_D$, the velocity terms $\sigma_v$ cancel out, leaving the
simple relation $\nabla \phi=-\sigma_\phi/\xi$, that relates the
stratification to the concentration fluctuations. To evaluate
$\sigma_\phi$, we use the same Poisson model as Mucha and Brenner,
$\sigma_\phi=\sqrt{\phi S(\phi) a^3/ \xi^3}$. Here, $S(\phi)$ is the
structure factor of random hard spheres needed at moderate to high
concentrations to account for  excluded volume effects
\cite{nature}. Finally, we solve for the correlation length $\xi$
and obtain
  \begin{equation} \xi/a=c_0 [\phi
  S(\phi)]^{1/5}[-a\nabla\phi]^{-2/5}, \label{eq:bm2}
    \end{equation}
which is identical to the MB model Eq. (\ref{eq:bm}).

The new insight provided by this re-derivation of the MB model is
that it explicitly shows that  Eq. (\ref{eq:bm}) is contingent upon
{\it steady state} behavior, in which $\partial\langle
\phi\rangle_\tau/\partial t=0$. This highlights the advantage of
testing this model in fluidized beds, which produce time invariant
concentration profiles. In sedimentation on the other hand, several
studies have reported that the concentration profile continues to
evolve as the column is falling\cite{shang,evolution,maxey}.
Consequently, the profile may or may not reach a steady state before
the column collapses to the bottom. This derivation explicitly shows
that in columns with time varying concentration profiles, i.e.
$\partial\phi(z,t)/\partial t\neq0$, the advective and diffusive
particle flux terms would not be balanced,  and the MB model Eq.
(\ref{eq:bm}) would not hold true.

\section{Discussion and Conclusions}\label{sec:conc}
In this article we present  results on the particle dynamics in
liquid fluidized beds at low Reynolds number. Our primary aim is to
examine the behavior of the fluctuating particle velocities and the
stationary concentration profiles. We chose to do this in moderately
concentrated samples, all of mean concentration $\phi_0=0.10$. To
examine how the velocity and concentration profiles depend upon
system size, we used a single cell geometry, but varied the particle
size $a$ by a factor of $\sim 4.5$ so that the system size $N$
varies by almost two orders of magnitude.

For all particle sizes, the measured concentration gradients
      $\nabla \phi$, fluctuation magnitudes $\sigma_v$, and
      correlation lengths $\xi$, all display $z-$dependent profiles
    in the particle columns. These profiles are found to change in
    magnitude, but not form, when the particle sizes $a$  are
    varied over a wide range. Specifically, we find a scaling
     regime for particles larger than a critical size, $a\gg a^*$,
     in which the data profiles from different particle sizes
     collapse together when plotted in the forms $-a\nabla \phi(z)$,
     $\xi(z)/a$, and $\sigma_v(z)/v_p$. (see Figs. \ref{fig:mids}(b),
      \ref{fig:dv}, and \ref{fig:xi}). These simple scalings with
      particle size demonstrate the universality of the particle
      dynamics and structure  in fluidization.

    The new scalings also quantify how the bed properties depend
     upon  system size. In our experiments, the  number of
     particles fluidized, $N=TWH\phi_0/(4/3)\pi a^3\propto 1/a^3$,
     increases with decreasing particle size. The scaling region is
      found when $a\gg a^*$, or equivalently $N\ll N^*$, where $N^*\approx 35$
      million particles. In this regime, the stratification scaling of
      $a\nabla \phi(z)$ can similarly be written as a $N^{-1/3}\nabla \phi(z)$ scaling,
      explicitly showing that the concentration gradient  grows with $N$. Interestingly,
       in the limit of very small systems, the gradients vanish, i.e.
       $\lim_{N\rightarrow 0} \nabla \phi(z)=0$, resulting in what would be a       uniformly concentrated particle bed. This is an important point that shows that
       it would generally be easier to observe  concentration gradients in
       experiments, as opposed to simulations, since experimental system
       sizes are usually much larger. On the other hand, Ladd \cite{tony1}
        simulated the sedimentation of $\sim 32000$ particles (in a square cell geometry)
        at $\phi_0=0.13$ and, while looking across the middle of the column, found evidence
         for an extremely small gradient $-a\nabla \phi$ "{\it of order $10^{-5}$}". This result
         agrees well with our measurements shown in Fig. \ref{fig:mids}, where near mid-height,
         $-a\nabla \phi\approx 2-3\cdot 10^{-5}$.

    Evidence is also presented that the particle dynamics in fluidized beds
     can be closely connected to that seen in similar
    systems that are sedimenting downwards. We first demonstrate  this  for
     a typical sample in Fig. \ref{fig:pumponoff}, where the magnitude of
     the velocity fluctuations show no discernible difference between the
      two modes. To pursue this further, we apply the same column averaging
       methods to our data that has  been used in sedimentation experiments
        to extract a single value for $\sigma_v$ and $\xi$ in each of our
        fluidized beds. These values were found to be in good agreement
        with published findings in sedimentation, further connecting the
         two. It is known, however, that clear differences do exist between
    fluidization and sedimentation, particulary regarding whether or not the
     systems are in a steady state. Numerous studies\cite{shang,maxey,russian} beginning with     Guazzelli\cite{evolution}, have shown that the particle dynamics
      ($\sigma_v$, $\xi$, $\nabla \phi \ldots$) in sedimentation can
      be strongly time-dependent as the particles fall, in which case
       a steady state is not achieved. In fluidized systems, on the other hand,
        as the column stays suspended indefinitely, a steady state is eventually
         reached, as we find here.  We note that in our fluidization/sedimentation
          comparisons, we compared to sedimentation data
           thought to be in  near steady state conditions.

 A further, and central, finding in this paper is the
verification in Sec. \ref{sec:sa} of the Mucha and Brenner model for the velocity correlation
lengths $\xi(z)$. The model predicts that $\xi(z)$ can be calculated solely from the concentration profile $\phi(z)$ (and its derivative $\nabla \phi(z)$). We  measured  $\phi(z)$, and found that
  the MB model  predictions for $\xi(z)$ from Eq. (\ref{eq:bm}) were in very  good agreement with our correlation length measurements in all four beds, and at all heights within each bed.   The physical argument used by Mucha and Brenner is that, owing to buoyancy forces,  concentration gradients limit the maximal size over which particle velocities can be correlated, and this leads to the relation between $\xi$ and $\nabla \phi$ given in Eq. (\ref{eq:bm}). In sec. \ref{sec:advec}, we also show an alternative derivation of Eq. (\ref{eq:bm}), starting from a general advection-diffusion equation. The physical picture  we used is seemingly different from the MB line of reasoning, yet both yield the same relation between  $\xi$ and $\nabla \phi$. In our
 derivation, when the system is in a  steady state, i.e.
 $\partial\langle \phi(z,t)\rangle_\tau/\partial t=0$,
 the diffusive flux generated by  concentration gradients,
  $D\nabla \phi$, is counterbalanced by an advective flux
  due to the fluctuations, $\sigma_\phi \sigma_v$, which leads to Eq. (\ref{eq:bm}).

Several key questions still remain to be answered, starting with the observation that the MB relation Eq. (\ref{eq:bm}) requires a
knowledge of the stratification in $\nabla \phi$ for a determination
of the correlation length $\xi$, but does not allow for an {\it a
priori} calculation of either quantity.  Additionally, while we
 found there to be an overall  system size dependence for several
 properties in steady state sedimentation, we did not vary the
 different dimensions of the cell (TWH) independently, so it is not known whether some dimensions of the cell are more influential in this regard than others. Future experiments using different aspect ratio cells are being planned to help resolve these issues.

We thank Shang Tee and Tony Ladd for stressing the importance of
using nearly monodisperse beads, and Alex Levine, Dave Weitz and
Michael Brenner for many discussions. (corresponding author:
psegre@physics.emory.edu)

\end{document}